\documentclass[pdflatex,sn-mathphys-num]{sn-jnl}

\usepackage{graphicx}%
\usepackage{multirow}%
\usepackage{amsmath,amssymb,amsfonts,bm}%
\usepackage{amsthm}%
\usepackage{mathrsfs}%
\usepackage{anyfontsize} 
\usepackage{xcolor}%
\usepackage{textcomp}%
\usepackage{booktabs}%
\usepackage{algorithm}%
\usepackage{algorithmicx}%
\usepackage{algpseudocode}%
\usepackage{listings}%
\usepackage{tikz,tikz-cd}

\usepackage{xr}
\usepackage{sidecap}

\newcommand{\comment}[1]{\textcolor{red}{#1}}
\newcommand{\unit}[1]{\mathrm{#1}}

\begin{document}

\title[Bonded-particle model for magneto-elastic rods]{Bonded-particle model for magneto-elastic rods}
\author[1,2]{\fnm{Gabriel} \sur{Alkuino}}\email{gsalkuin@syr.edu}
\author*[3]{\fnm{Joel T.}~\sur{Clemmer}}\email{jtclemm@sandia.gov}
\author[1,2]{\fnm{Christian D.}~\sur{Santangelo}}\email{cdsantan@syr.edu}
\author*[2,4]{\fnm{Teng} \sur{Zhang}}\email{tzhang48@syr.edu}

\affil[1]{\orgdiv{Department of Physics}, \orgname{Syracuse University}, \orgaddress{\city{Syracuse}, \state{NY}, \postcode{13244}, \country{United States}}}
\affil[2]{\orgdiv{BioInspired Syracuse}, \orgname{Syracuse University}, \orgaddress{\city{Syracuse}, \state{NY}, \postcode{13244}, \country{United States}}}
\affil[3]{\orgname{Sandia National Laboratories}, \orgaddress{\city{Albuquerque}, \state{NM}, \postcode{87123}, \country{United States}}}
\affil[4]{\orgdiv{Department of Mechanical and Aerospace Engineering}, \orgname{Syracuse University}, \orgaddress{\city{Syracuse}, \state{NY}, \postcode{13244}, \country{United States}}}

\abstract{
We develop a bonded-particle model for magneto-elastic rods that unifies large deformations, contact, and long-range magnetic interactions within a single discrete-element framework.
The rod is discretized into orientable particles connected by co-rotational bonds that capture stretching, shearing, twisting, and bending through a symmetric decomposition of relative displacement and rotation.
Magnetic coupling is introduced at the particle level: each particle carries a dipole moment that rotates with it, enabling both external-field actuation and long-range dipole--dipole interactions without modifying the structural formulation.
We implement the model in LAMMPS to take advantage of its parallel efficiency, long-range electrostatic solvers, and multiphysics capabilities.
We validate the model on three problems spanning writhing instabilities, non-uniform magnetic actuation, and dipole-induced mechanical hysteresis.
To demonstrate multiphysics capability, we couple the model with a lattice Boltzmann fluid solver via the immersed boundary method and simulate filaments in oscillatory channel flow and fluid pumping by magnetically actuated cilia arrays.
Across all examples, the model shows good agreement with experimental, analytical, and numerical reference results.
}

\maketitle

\section*{Introduction}\label{sec:intro}

Hard-magnetic soft materials have emerged as a promising platform for untethered soft robots, as they can be programmed and remotely controlled via external magnetic fields \citep{lum_shape-programmable_2016,Hu2018}. 
This has opened new possibilities in microrobotics for biomedical applications \citep{peyer2013bio,yang2020magnetic}. 
Additionally, magnetic dipole--dipole interactions can be harnessed to induce multistability \citep{gu2023self,korpas2021temperature} and mechanical hysteresis \citep{liang2022phase,sano2022kirchhoff,sun2026kirigami}, and form self-assembling network structures \citep{gu2019quadrupole,yang2023_networks}, expanding the metamaterial design space \citep{Aghamiri2025}. 
In numerous applications, the fundamental building blocks are \textit{beams} or \textit{rods}, owing to their capacity for large deformations and complex shape morphing. 
They have been used to design bio-inspired swimmers \citep{peyer2013bio}, robots \citep{lee2023magnetically}, magneto-mechanical metamaterials \citep{zou2023magneto}, surgical catheters \citep{hwang2020review,dreyfus2024dexterous}, and artificial cilia \citep{Gu2020,park2023bioinspired}.
Accurately and efficiently simulating such systems in complex environments is challenging due to the interplay of large deformations, nonlocal interactions, contact, and other external forces. 

Several discrete rod formulations exist, including those based on co-rotational finite elements~\citep{crisfield1990consistent,le2014consistent,yang20223d}, Cosserat rod theory and discrete differential geometry~\citep{bergou2008discrete,gazzola2018forward,huang2023discrete}, and bonded-particle models~\citep{potyondy2004bonded,carmona2008fragmentation,andre2012discrete,chen_comparative_2022,zhang2024rod}.
For hard-magnetic rods, reduced-order rod models augmented with the Zeeman energy offer efficient alternatives to three-dimensional finite-element models \citep{yan_comprehensive_2022,sano2022kirchhoff,yang20223d,huang2023discrete,yashraj_bhosale_2023_7658892}. 
These rod models have been extended to include dipole--dipole interactions \citep{sano2022reduced,huang2023discrete} and have been implemented in various simulation frameworks, including those capable of fluid--structure interaction \citep{tekinalp2025self,huang2025tutorial}. 

In geomechanics, bonded-particle models are widely used to simulate elasticity and fracture in 2D and 3D solids \citep{potyondy2004bonded,lisjak2014review,potyondy2015bonded,chen_comparative_2022}. 
Yet their application to 1D structures undergoing extreme deformation remains relatively unexplored, despite their natural suitability for soft matter systems across scales---from polymers to mechanical metamaterials. 
Indeed, recent works have applied bonded-particle models to liquid crystals \citep{hackney2026shape}, entangled matter \citep{pezeshki2025tunable}, and elasto-granular systems \citep{pol2025granular}, highlighting their growing adoption beyond traditional geomechanics.
Although several studies have applied bonded-particle models to simulate rods and fibers \citep{guo2013validation,nguyen2013validation,chen_comparative_2022}, they remain limited to pure planar bending or twisting cases and their stability remains unproven under scenarios of extreme deformation.
Moreover, their application to magneto-elastic materials remains largely unexplored.

While the term bonded-particle model originates from \citet{potyondy2004bonded}, we use it here broadly to refer to any non-contact bonded interaction for modeling solid behavior. 
The simplest models treat a solid as a network of simple springs \citep{nealen2006physically}; these are sometimes called mass-spring models (MSM) or lattice-spring models. 
\citet{lloyd2007identification} optimized the MSM for triangular (2D) and tetrahedral (3D) meshes and obtained good agreement with finite element method (FEM) results, though the Poisson's ratio cannot be independently specified ($\nu_\text{2D} = 1/3$ \citep{seung1988defects} and $\nu_\text{3D} = 1/4$ \citep{greaves2013poisson}). 
Furthermore, they found that, in 3D, a volume correction term is needed for better accuracy. 
\citet{clemmer2024soft} used an MSM based on random sphere packing and introduced a deviatoric term---computed from the difference between local volumetric dilation and bond stretching---to access a larger range of Poisson's ratios. 
\citet{golec2020hybrid} and \citet{zhang2019deriving} derived mass-spring equivalents of hexahedral mesh elements for linear isotropic and neo-Hookean solids, respectively, with both models including a multi-body volume correction term.

\citet{ye2021magttice} extended the model of \citet{zhang2019deriving} to include magnetic actuation.
However, adding dipole--dipole interactions is not straightforward since the nodes are point particles. 
More generally, purely radial springs are insufficient for capturing the coupling between shear and rotation \citep{wang_implementation_2006}. 
Some bonded-particle models address these limitations by incorporating shear and rotational springs through finite-sized, orientable particles. 
This enables dipoles to be assigned directly to the discrete elements. 
However, the added rotational degrees of freedom make it difficult to derive the correct spring parameters from macroscopic elasticity in 2D and 3D \citep{wang2008macroscopic}.

For 1D systems, many bonded-particle models based on rod theory exist \citep{carmona2008fragmentation,andre2012discrete,obermayr2013bonded, zhang2024rod}. 
\citet{chen_comparative_2022} reviewed various bonded-particle models and unified rod-based bond models into a generic form. 
In this work, we develop a bonded-particle model (BPM) for magneto-elastic rods (Fig.~\ref{fig:magnetic-bpm-intro}(a)) that integrates seamlessly with discrete element method (DEM) and molecular dynamics frameworks, offering a unified treatment of large deformations, contact, and long-range magnetic interactions. 
We implement our model in LAMMPS \citep{thompson2022lammps}, leveraging its robust infrastructure for large-scale systems, long-range electrostatics, and multiphysics coupling. 

Furthermore, many applications of magneto-elastic rods involve operation in a fluid environment, where hydrodynamic forces shape the system's dynamics and vice versa---from fluid-pumping artificial cilia to bio-inspired swimming robots \citep{peyer2013bio,yang2020magnetic,hwang2020review,park2023bioinspired}.
In biophysics and microfluidics, the immersed boundary method (IBM) coupled with the lattice Boltzmann method (LBM) is a well-established approach for simulating fluid--structure interaction \citep{kruger2017lattice}.
Recent works include coupling a Cosserat rod model with an isogeometric IBM \citep{agrawal2024efficient} and with a velocity--vorticity Navier--Stokes formulation \citep{tekinalp2025self}.
To this end, we also develop a lightweight LBM--IBM package in LAMMPS and couple it with our rod model.

\section*{Results}

\subsection*{Bonded-particle model for elastic rods}\label{sec:BPM-elastic}

\begin{figure}[!htpb]
    \centering
    \includegraphics[width=\linewidth]{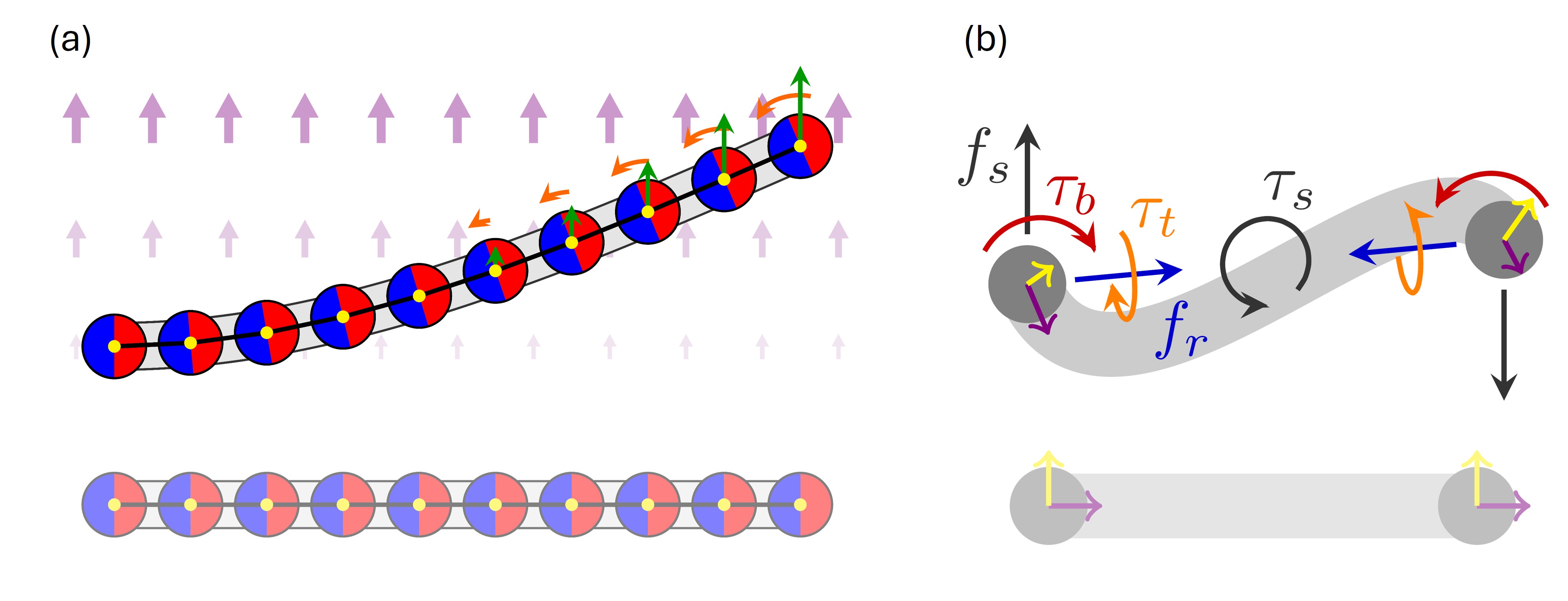}
    \caption{Illustration of the bonded-particle model for magneto-elastic rods. (a)~Each particle is treated as a magnetic dipole. An external magnetic field or dipole--dipole interaction exerts a force and torque on the particle. Neighboring particles are connected by an elastic bond. (b)~The relative displacement and rotation between two bonded particles are decomposed into four linear spring interactions. The reference configurations are displayed below the corresponding deformed configurations.}
    \label{fig:magnetic-bpm-intro}
\end{figure}

We build upon the BPM formulation of Wang et al.~\citep{wang_implementation_2006,wang2008macroscopic,wang2009new, wang_esys_particle_2009}.
Their rotation decomposition technique extracts sequence-independent bending and twisting angles directly from the reference and current configurations, avoiding the instabilities of incremental formulations \citep{wang2009new}.
Each cylindrical bond element has four linear springs corresponding to stretching, shearing, twisting, and bending (Fig.~\ref{fig:magnetic-bpm-intro}(b)). 
The elastic forces and torques are linearized functions of relative displacement and rotation, making the BPM a co-rotational model. 
Our formulation differs from Wang et al.~\citep{wang_implementation_2006,wang2008macroscopic,wang2009new,wang_esys_particle_2009} in that we do not use one particle as the reference frame. 
Following other co-rotational formulations in mechanics \citep{simo1985finite,rankin1986element,crisfield1990consistent,felippa2005unified,le2014consistent,cottanceau2018finite,grange2024co}, we introduce a \textit{central frame} obtained by averaging the two particles' orientations via quaternion spherical linear interpolation, similar to \citet{obermayr2013bonded}. 
The radial and shear forces arise as restoring forces from displacement components along and transverse to the current bond direction, respectively. 
The central frame eliminates the rotational shear term and ensures the formulation is symmetric under particle exchange (Supplementary Information~\ref{S-sec:moreBPM}).

\begin{figure}[!htpb]
    \centering
    \includegraphics[width=\linewidth]{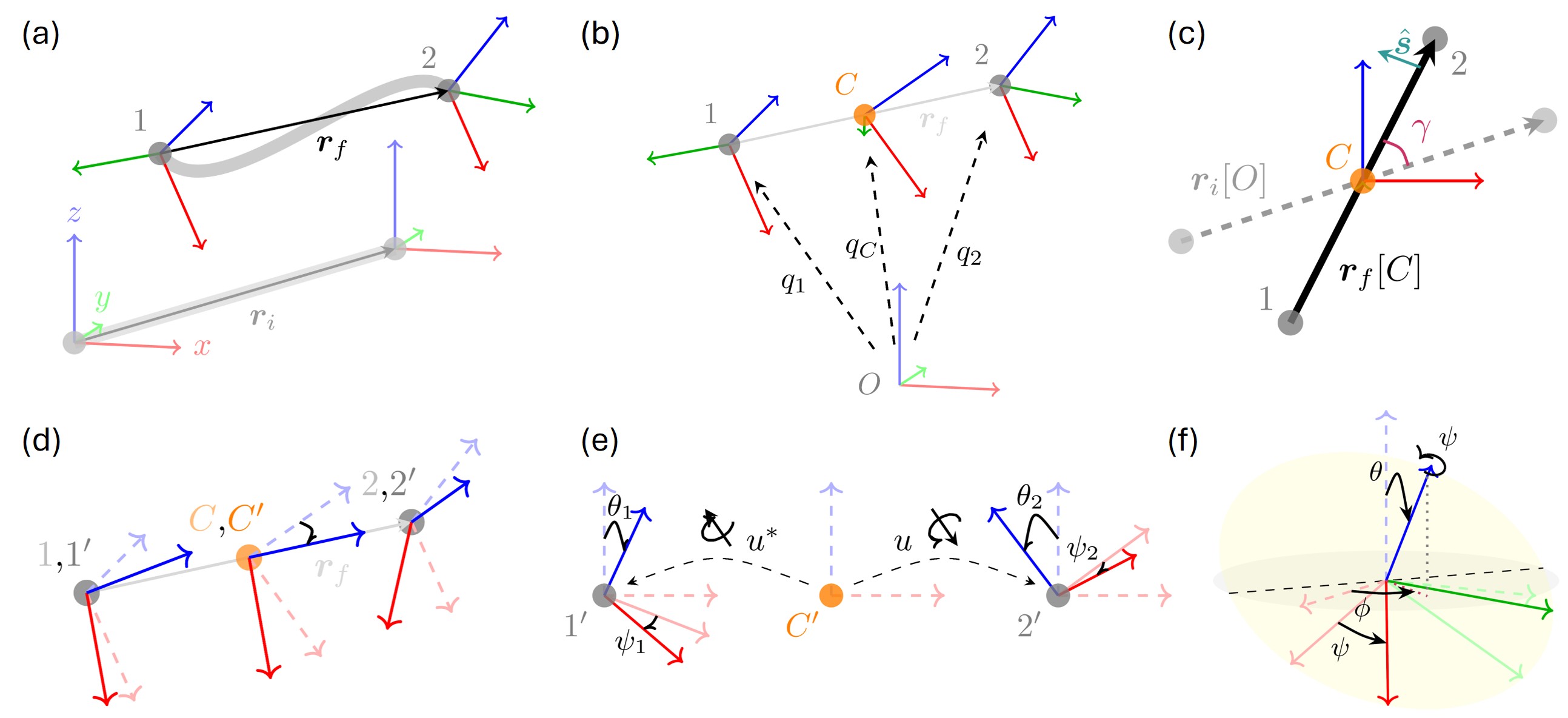}
    \caption{Definition of the BPM frames. (a)~The reference (light) and deformed (dark) bond. (b)~The body frames of the two particles are defined by the quaternions $q_1$ and $q_2$ in the global frame. The central frame, $C$, is defined by $q_C$, which is the average of $q_1$ and $q_2$. (c)~The relative displacement is measured from the moving $C$ frame. (d)~The $C'$ frame is obtained by aligning the $C$ frame to the bond vector. (e)~The orientation of particles $1$ and $2$ in the $C'$ frame is defined by the quaternions $u^\ast$ and $u$, respectively. Both quaternions are then decomposed in a \textit{bend before twist} order such that the total rotation of particle $2$ with respect to particle $1$ is a sequence of four rotations. (f)~Illustration of the swing--twist decomposition.}
    \label{fig:BPM-frames}
\end{figure}

Suppose a reference bond is initialized between two particles $1$ and $2$ with position and orientation (represented by a unit quaternion) defined by the tuples $(\vec{x}_1^{\,i}, q_1^{\,i})$ and $(\vec{x}_2^{\,i}, q_2^{\,i})$, respectively. 
Without loss of generality, let $q_1^{\,i} = q_2^{\,i} = 1$. 
In the \textit{current} deformed configuration, the particles are described by $(\vec{x}_1^{\,f}, q_1^{\,f})$ and $(\vec{x}_2^{\,f}, q_2^{\,f})$. 
We decompose the rigid-body motion taking particle $1$ to $2$ into four distinct translational or rotational modes, each associated with a linear spring of stiffness we call $K_r$, $K_s$, $K_t$, or $K_b$. 
The initial and current bond vectors are $\bm{r}_i := \vec{x}_2^{\,i} - \vec{x}_1^{\,i}$ and $\bm{r}_f := \vec{x}_2^{\,f} - \vec{x}_1^{\,f}$, respectively. 
We define the central frame $C$ with origin at the bond midpoint and orientation $q_C$ given by the average of the two particle orientations, where $q_C^{\,i} = 1$. 
During the deformation, the central frame evolves from $C^i$ to $C^f$. 
Therefore, the relative displacement must be evaluated in the moving $C$ frame (Fig.~\ref{fig:BPM-frames}(a)-(c)):
\begin{equation}
    \Delta \bm{r}[C^f] = \bm{r}_f[C^f] - \bm{r}_i[C^i] = q_C^\ast \bm{r}_f[O] \, q_C - \bm{r}_i[O].
\end{equation}
From here on, since the initial frames are the same as the global frame $O$, the frames $1$, $2$, and $C$ without the $f$ superscript refer to the current configuration. 
For small relative displacements, the axial component of $\Delta \bm{r}$ is approximately $(r_f - r_i) \hat{\bm{r}}_f$. 
Therefore, the radial (axial) force on particle $1$ is
\begin{equation}\label{eq:bpm-fr}
    \bm{f}_r := K_r (r_f - r_i) \hat{\bm{r}}_f.
\end{equation}
The shear force acts perpendicular to $\hat{\bm{r}}_f$ and lies in the plane spanned by $\hat{\bm{r}}_i[O]$ and $\hat{\bm{r}}_f[C]$. 
Let $\gamma = \arccos{(\hat{\bm{r}}_{i}[O] \cdot \hat{\bm{r}}_f[C])}$ and define two unit vectors $\hat{\bm{t}}[C] = (\hat{\bm{r}}_i[O] \times \hat{\bm{r}}_f[C]) / \big| \hat{\bm{r}}_i[O] \times \hat{\bm{r}}_f[C] \big|$ and $\hat{\bm{s}} = \hat{\bm{t}} \times \hat{\bm{r}}_f$, such that $(\hat{\bm{s}}, \hat{\bm{t}}, \hat{\bm{r}}_f)$ form a right-handed basis. 
The shear force on particle $1$ due to the transverse displacement is then
\begin{equation}\label{eq:bpm-fst}
    \bm{f}_s := K_s \, r_f \, \gamma \, \hat{\bm{s}}.
\end{equation}
This force also produces a counter-torque to conserve angular momentum:
\begin{equation}\label{eq:bpm-tst}
    \bm{\tau}_s := \frac{1}{2} \bm{r}_f \times \bm{f}_s = \frac{1}{2} K_s \, r_f^2 \, \gamma \, \hat{\bm{t}}.
\end{equation}
The forces and torques are conveniently computed in $C$ and then rotated back to the global frame via $\bm{v}[O] = q_C \, \bm{v}[C] \, q_C^\ast$.

To extract the bending and twisting angles, we define the $C'$ frame by rotating $C$ such that its $z$-axis aligns with the bond vector (Fig.~\ref{fig:BPM-frames}(d)); the rotation is given by the quaternion $m$ (Eq.~\eqref{S-eq:quat-h}) in the $C$ frame. 
In the $C'$ frame, each particle's orientation is decomposed via swing--twist decomposition (Fig.~\ref{fig:BPM-frames}(f)), yielding (half) bending rotations $\mathbf{R}(\theta_\alpha, \hat{\bm{z}} \times \hat{\bm{\phi}}_\alpha)$ and (half) twisting rotations $\mathbf{R}(\psi_\alpha, \hat{\bm{z}})$ for $\alpha = 1 \text{ and } 2$, all defined in $C'$ (Fig.~\ref{fig:BPM-frames}(e)), where $\mathbf{R}(\Phi,\hat{\bm{n}})$ is the rotation matrix that rotates by the angle $\Phi$ about the axis $\hat{\bm{n}}$. 
This can be done using Wang's \citep{wang2009new} algorithm or other swing--twist decomposition algorithms \citep{swing-twist}.
The effective bending and twisting rotations are then defined as
\begin{align}
    \mathbf{R}_b &= \mathbf{R}(\theta_2, \hat{\bm{z}} \times \hat{\bm{\phi}}_2) \, \mathbf{R}^{-1}(\theta_1, \hat{\bm{z}} \times \hat{\bm{\phi}}_1), \\
    \mathbf{R}_t &= \mathbf{R}(\psi_2 - \psi_1, \hat{\bm{z}}).
\end{align}
The bending and twisting torques on particle $1$ are
\begin{align}
    \bm{\tau}_b[C'] &:= K_b \bm{\theta} = K_b \theta \hat{\bm{b}}, \label{eq:bpm-tb} \\
    \bm{\tau}_t[C'] &:= K_t \bm{\psi} = K_t (\psi_2 - \psi_1) \hat{\bm{z}}, \label{eq:bpm-tt}
\end{align}
where $\bm{\theta} = \theta \hat{\bm{b}}$ and $\bm{\psi} = (\psi_2 - \psi_1) \hat{\bm{z}}$ are the axis--angle vectors of $\mathbf{R}_b$ and $\mathbf{R}_t$, respectively. 
These are then rotated back to the global frame via $\bm{v}[O] = q_C \, m \, \bm{v}[C'] \, m^\ast q_C^\ast$. 
Finally, the net force and torque on each particle in the global frame are obtained by linear superposition:
\begin{equation}\label{eq:bpm-vector}
    \begin{bmatrix}
    \bm{f}_1 \\
    \bm{\tau}_1 \\
    \bm{f}_2 \\
    \bm{\tau}_2 
    \end{bmatrix} =
    \begin{bmatrix}
    \bm{f}_r + \bm{f}_s \\
    \bm{\tau}_s + \bm{\tau}_b + \bm{\tau}_t \\
    -\bm{f}_r - \bm{f}_s \\
    \bm{\tau}_s - \bm{\tau}_b - \bm{\tau}_t
    \end{bmatrix}.
\end{equation}

The spring constants consistent with Euler--Bernoulli beam theory are $K_r = EA/L$ (stretch), $K_s = 12EI/L^3$ (``shear''), $K_t = GJ/L$ (twist), and $K_b = EI/L$ (bend), where the bond is assumed to be a cylinder of length $L$ and diameter $d$, with cross-sectional area $A = \pi d^2/4$, planar second moment of area $I = \pi d^4/64$, and polar second moment of area $J = \pi d^4/32$. 
We note that the term ``shear'' here refers to the transverse displacement in the central frame, and not to a physical shear deformation as in Timoshenko beam theory.
The ``shear'' stiffness $K_s$ is the transverse stiffness of an Euler--Bernoulli beam element (Supplementary Information \ref{S-sec:bpm-constants}).
Although $K_s$ is the stiffest of the four springs, making transverse displacements the most energetically unfavorable mode, the term is necessary to reproduce the full Euler--Bernoulli stiffness matrix.
This formulation is consistent with the unified bond model framework of \citet{chen_comparative_2022}.

The total energy of the bond is approximately 
\begin{equation}\label{eq:total-energy}
    U \approx \frac{1}{2} K_r (r_f-r_i)^2 + \frac{1}{2} K_s r_f^2 \gamma^2 + \frac{1}{2} K_t \psi^2 + \frac{1}{2} K_b \theta^2.
\end{equation}
This is a Cosserat-like discrete energy with stretching, ``shearing'', twisting, and bending contributions. 
We implement this model as a custom bond style in the LAMMPS BPM package \citep{clemmer2024soft}.
The equations of motion are integrated using velocity Verlet for translations and angular velocities, and Richardson extrapolation for quaternion orientations (Supplementary Information \ref{S-sec:BPM-pkg}).

\subsection*{Extreme twisting and plectoneme formation}\label{sec:heavy-rods}

The writhing instabilities and plectoneme formation in twisted elastic rods have been extensively studied, both analytically and experimentally \citep{coyne1990analysis,thompson1996helix,goriely1998nonlinear,goss2005experiments}, and have served as benchmark problems for discrete rod models \citep{bergou2008discrete,gazzola2018forward,cottanceau2018finite,grange2024co}. 
Here, we apply our model to simulate the twisting of heavy elastic rods, initially compressed such that they buckle under their own weight, as studied by \citet{lazarus2013continuation,lazarus2013contorting} and further investigated by \citet{cottanceau2018finite,grange2024co}. 
Two cases are considered: a straight rod and a naturally curved rod.

\begin{figure}[!htpb]
    \centering
    \includegraphics[width=\linewidth]{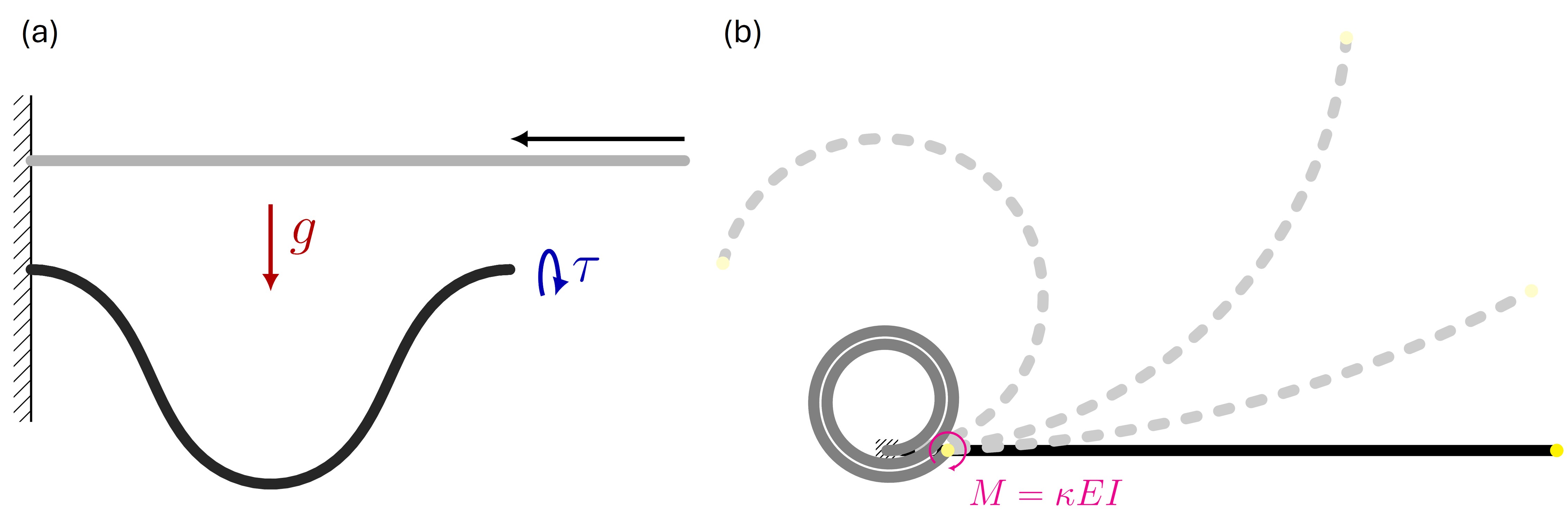}
    \caption{The setup for the twisting experiment of \citet{lazarus2013continuation}. (a)~An initially straight rod is compressed by a fixed amount causing it to buckle under its own weight. The rod is then twisted multiple times and the resulting shape is studied (Fig.~\ref{fig:heavy-straight-rod}). (b)~An ideal naturally curved rod coiled into a circle with curvature $\kappa$ can be straightened by applying a pure end moment $M = \kappa E I $. The straightened rod then undergoes the same twisting procedure as in (a) (Fig.~\ref{fig:heavy-curved-rod}).}
    \label{fig:heavy-rods-setup}
\end{figure}

For the first example, a straight rod of length $L = 300~\unit{mm}$, diameter $d = 3.1~\unit{mm}$, density $\rho = 1.2~\unit{g}/\unit{cm}^3$, and elastic moduli $E = 1.3~\unit{MPa}$ and $G = 433~\unit{kPa}$ was discretized into $N=300$ bonds. 
In the simulation, the leftmost particle was excluded from time-integration such that its position and orientation were fixed, corresponding to a clamped boundary condition. 
The simulation was performed in two stages (Fig.~\ref{fig:heavy-rods-setup}(a)).
First, the rod was axially compressed by $80~\unit{mm}$ by displacing the rightmost particle at constant velocity while keeping its orientation fixed, and then holding the particle fixed for some duration to allow the rod to equilibrate.
Second, the rod was twisted up to $9$ full rotations by rotating the same particle about the initial rod axis at a constant rate. 

\begin{figure}[!htpb]
    \centering
    \includegraphics[width=\linewidth]{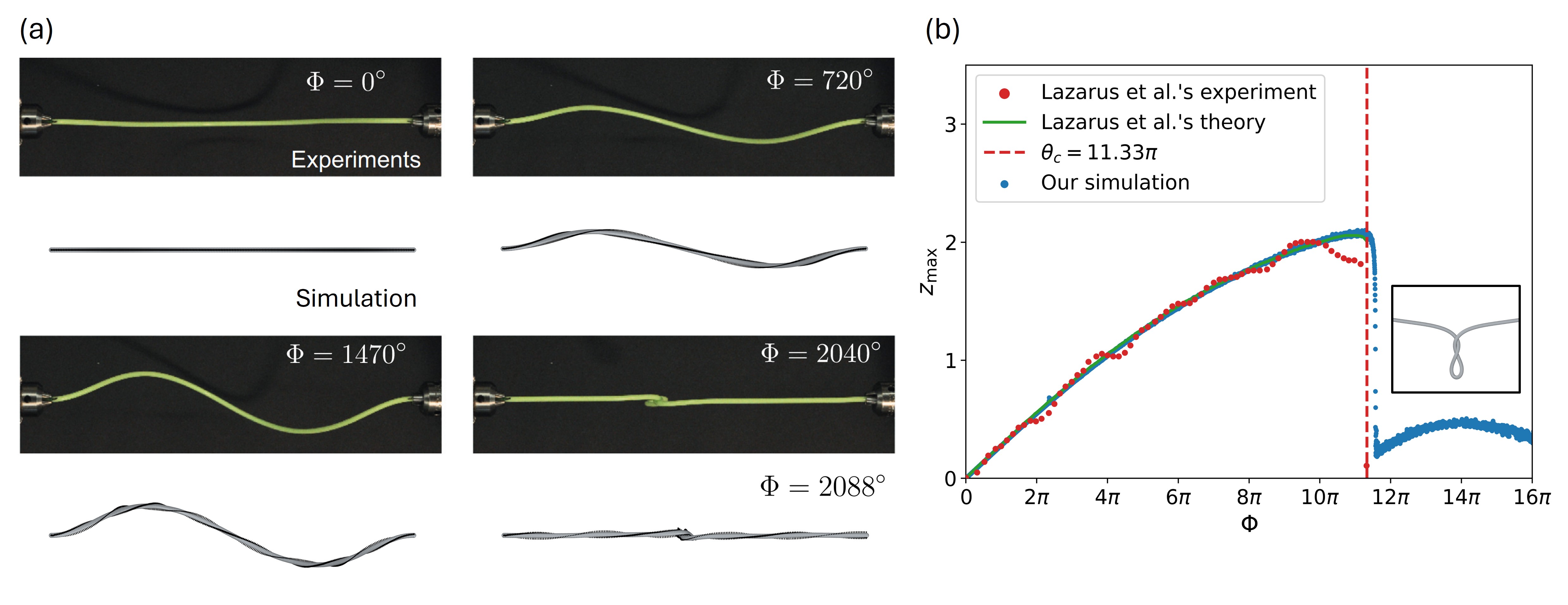}
    \caption{Extreme twisting of a pre-buckled initially straight rod. (a)~Top view snapshots at different twisting angles $\Phi$ comparing the experimental images of \citet{lazarus2013continuation} and ours. (b)~The maximum deflection as a function of the twist angle. The red points are their experimental measurements and the green curve is their semi-analytical solution, while the blue curve is our simulation result. The simulated plectoneme (inset) formation happens near their predicted critical angle $\theta_c = 11.33\pi$.}
    \label{fig:heavy-straight-rod}
\end{figure}

Fig.~\ref{fig:heavy-straight-rod} compares our simulation results with Lazarus et al.'s \citep{lazarus2013continuation}. 
Simulation snapshots were taken every $3.6^\circ$, and the representative images used for comparison in Fig.~\ref{fig:heavy-straight-rod}(a)~were those nearest to the listed experimental angles. 
The images show the \textit{top view}, where gravity points into the page.
The twisting simulation is shown in Supplementary Video~1.
Fig.~\ref{fig:heavy-straight-rod}(b)~plots the maximum deflection, out of the plane of the initial buckled rod, as the rod is twisted, compared against their semi-analytical solution. 
The simulated deflection is in excellent agreement with their theory and experiment. 
The instability occurs slightly past the predicted critical angle of $\theta_c = 11.33\pi$, beyond which a plectoneme forms (Fig.~\ref{fig:heavy-straight-rod}(b)~inset). 
This delay is expected because finite loading rates allow the system to persist on the stable branch beyond the theoretical bifurcation point due to inertial effects. 
Additionally, our physics-based simulations can only access the stable branch of their solution. 
This contrasts with continuation methods, which can trace both stable and unstable branches by solving the equilibrium equations directly. 

For the second example, a rod with the same parameters was used, except this time it was initially curved. 
To prepare the curved rod of length $L$, particles were placed along $\vec{x}_i = R \langle \cos{(\theta_i-\pi/2)}, 1+\sin{(\theta_i-\pi/2)}, 0 \rangle$, where $\theta_i = iL/(NR), \; i = 0,1,\ldots,N$. 
The curvature of the rod is $\kappa = 1/R = 44.84~\unit{m}^{-1}$, causing it to coil into a circle $2.14$ times. 
In the simulations, this should not pose a problem unless contact is considered or the particles exactly overlap. 
Bonds were initialized in the coiled configuration. 
The curved rod was first straightened by slowly applying a moment $M = \kappa E I$ to the free end (Fig.~\ref{fig:heavy-rods-setup}(b)), following \citet{grange2024co}.
Gravitational and contact forces were excluded during this stage.
The simulation successfully captures the straightening of the coiled rod, providing validation that the model works for naturally curved rods without additional modification.
Once straightened, the rod was compressed and twisted following the same procedure as in the straight rod case (Fig.~\ref{fig:heavy-rods-setup}(a)), with walls added at both ends to prevent the plectoneme loop from slipping off.
The twisting simulation is shown in Supplementary Video~2.
Fig.~\ref{fig:heavy-curved-rod} summarizes the results, and again we observe excellent agreement with \citet{lazarus2013continuation}.

\begin{figure}[!htpb]
    \centering
    \includegraphics[width=\linewidth]{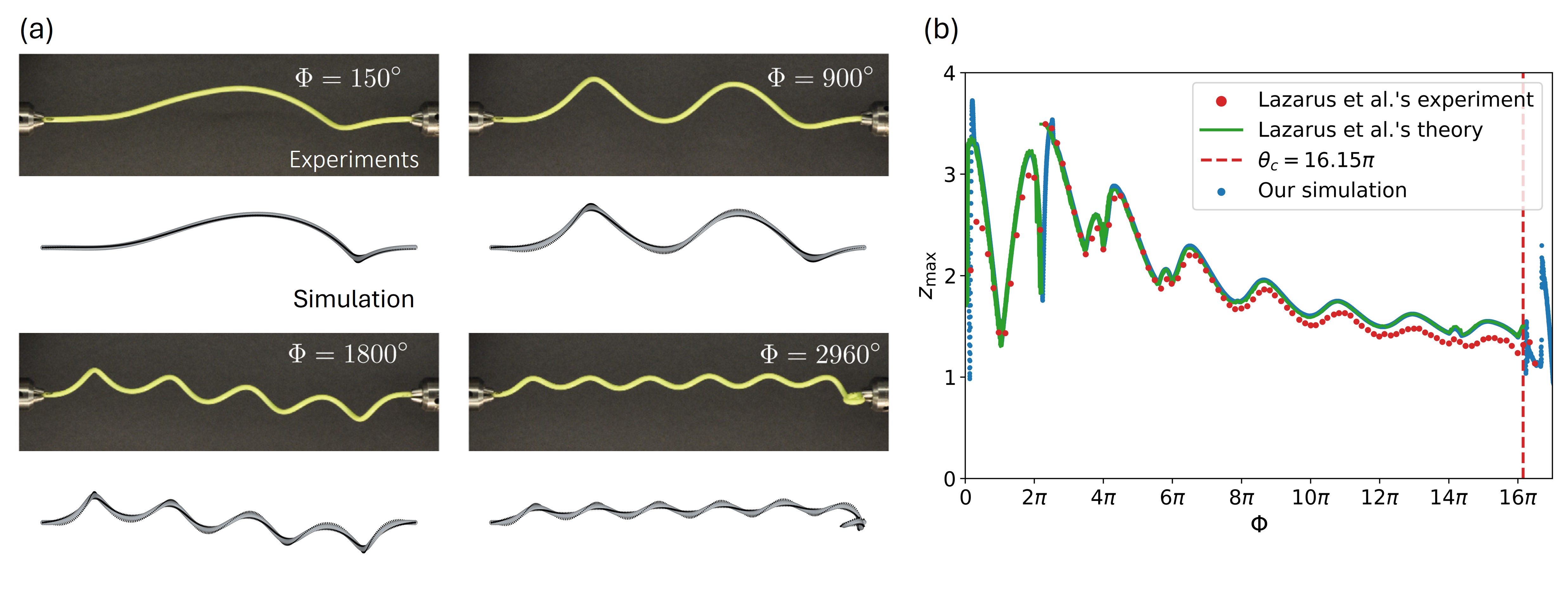}
    \caption{Extreme twisting of a pre-buckled, naturally curved rod. The rod is coiled in its stress-free state and straightened before undergoing the same procedure as in Fig.~\ref{fig:heavy-straight-rod}. (a)~Top view snapshots at different twisting angles comparing the experimental images of \citet{lazarus2013continuation} with ours. (b)~Maximum deflection as a function of twist angle. The red points are their experimental measurements and the green curve is their semi-analytical solution, while the blue curve is our simulation result. The plectoneme forms at a critical angle $\theta_c = 16.15 \pi$.}
    \label{fig:heavy-curved-rod}
\end{figure}

\subsection*{Magnetized beams in non-uniform fields}\label{sec:magnetized-beams}

Various constitutive models for hard-magnetorheological elastomers (hMRE) have been developed by the solid mechanics community \citep{dorfmann2014nonlinear,lu2024mechanics}. 
For hard-magnetic slender rods with minimal stretching, simple phenomenological models have gained widespread adoption owing to their practicality \citep{lum_shape-programmable_2016,wang2020hard}. 
Here, we use the following model for ideal hMREs \citep{lum_shape-programmable_2016,danas_stretch-independent_2024}
\begin{equation}
    \bm{M} = J^{-1} \mathbf{R} \tilde{\bm{M}},
\end{equation}
where $\tilde{\bm{M}}$ and $\bm{M}$ are the magnetization vectors in the reference and current configurations, respectively, $\mathbf{R}$ is the rotational component of the deformation gradient $\mathbf{F}$, and $J = \det{\mathbf{F}}$. 
Therefore, a discrete dipole element transforms as 
\begin{equation}
    \tilde{\bm{m}} = \tilde{\bm{M}} \tilde{V} \to \bm{m} = \bm{M} V = \left( J^{-1} \mathbf{R} \tilde{\bm{M}} \right) \left( J \tilde{V} \right) = \mathbf{R} \tilde{\bm{m}}.
\end{equation}
In other words, the dipole moment simply rotates with the particle (Fig.~\ref{fig:magnetic-bpm-intro}(a)). 
Assigning dipole moments directly to particles rather than bonds makes the model compatible with existing particle simulators.

We validate the magneto-elastic model against the comprehensive theoretical and experimental work of \citet{yan_comprehensive_2022} on beam bending using uniform and constant-gradient magnetic fields. 
They developed a 1D variational model, which they showed to be equivalent to an earlier model by \citet{lum_shape-programmable_2016}. 
Both models assume an initially straight beam in 2D and yield a second-order integro-differential equation. 
This equation can be difficult to solve for more complex magnetization profiles or external fields---such as those from portable magnets. 
The BPM sidesteps this since it computes the force and torque on each particle locally.

\begin{figure}[!htpb]
    \centering
    \includegraphics[width=0.8\textwidth]{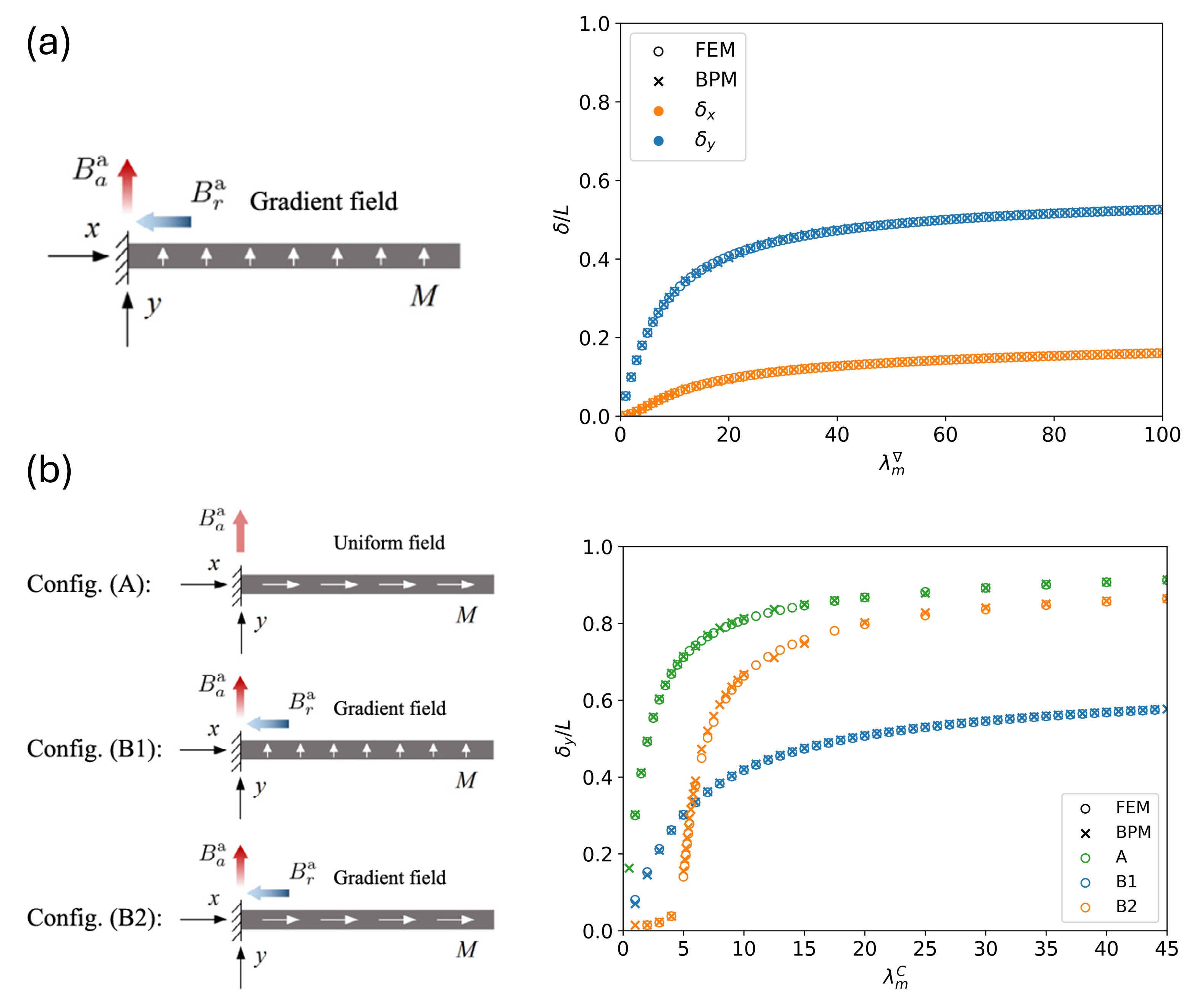}
    \caption{Magnetized beams under uniform and constant-gradient fields. The schematics in the left column were taken from \citet{yan_comprehensive_2022}. (a)~The beam is magnetized in the $+y$ direction and $(\nabla \bm{B})_{yy} > 0$. The $x$ and $y$ components of the tip displacement are compared with FEM results. (b)~The tip deflection in the $y$ direction compared with FEM results for three different configurations.}
    \label{fig:magnetic-beams}
\end{figure}

We simulated some of the configurations in \citet{yan_comprehensive_2022}, shown in the left column of Fig.~\ref{fig:magnetic-beams}. 
All examples use a rectangular beam with geometry ($L = 25.8~\unit{mm}$, $w = 1.21~\unit{mm}$, $h = 0.49~\unit{mm}$), Young's modulus $E = 1.16~\unit{MPa}$, Poisson's ratio $\nu \approx 0.5$, and density $\rho = 2.01~\unit{g}/\unit{cm}^3$. 
Here, the beam can only bend in one plane and cannot twist; therefore, the relevant spring constants $K_r$, $K_s$, and $K_b$ can be consistently defined even for a non-cylindrical cross-section.
For a rectangular beam, $I = wh^3 / 12$.
The magnetization has magnitude $M = 94.1~\unit{kA}/\unit{m}$ and direction that varies by case. 
The beam was discretized into 258 bonds, each $0.1~\unit{mm}$ in length, with the leftmost particle clamped.

For the configuration in Fig.~\ref{fig:magnetic-beams}(a) (left), the beam was first relaxed under gravity (acting along $+x$) using damped dynamics until equilibrium was reached. 
A magnetic field was then applied incrementally via the scalar potential $\psi = -(b/\mu_0) \bigl[y^2/2 - ((x-L/2)^2 + z^2)/4\bigr]$, where $b$ is a variable prefactor. 
This potential yields the magnetic field $\bm{B} = -\mu_0 \nabla \psi = b \, \langle -(x-L/2)/2, y, -z/2 \rangle$ with gradient $\nabla\bm{B} = b \,\mathrm{diag}(-1/2, 1, -1/2)$, producing a torque and force on each particle. 
The dimensionless parameter $\lambda_m^\nabla := M b A L^3/(EI)$ was increased from $0$ to $100$ in steps, with the beam allowed to equilibrate at each increment using damped dynamics. 
Fig.~\ref{fig:magnetic-beams}(a)~plots the $x$ and $y$ components of the tip displacement as a function of $\lambda_m^\nabla$, showing excellent agreement with reference FEM results obtained using FEniCSx. 

Next, we compare the $y$-deflection of three configurations: A, B1, and B2, shown in Fig.~\ref{fig:magnetic-beams}(b) (left). 
The simulation setup is the same as before, except there is no gravity.
Configuration B1 is similar to Fig.~\ref{fig:magnetic-beams}(a), except that the field is centered at the fixed end of the beam, $(0,0,0)$, rather than at the beam center, $(L/2,0,0)$.
Both A and B2 have beams magnetized in the $+x$ direction.
A is subjected to a uniform field $\bm{B} = B\,\hat{y}$, while B2 is subjected to the same constant-gradient field as B1 with an additional $0.01~\unit{mT}$ in the $+y$ direction to kick the beam out of equilibrium. 
We define the dimensionless parameter $\lambda_m^C := M B A L^2/(EI)$, with $\lambda_m^\nabla \approx 0.36 \lambda_m^C$ for this particular experiment \citep{yan_comprehensive_2022}. 
Fig.~\ref{fig:magnetic-beams}(b)~plots the tip deflection in the $y$ direction as a function of $\lambda_m^C$, again showing excellent agreement with reference FEM results.

\subsection*{Helical rods with dipole--dipole interactions}\label{sec:magnetic-helix}

Having validated magnetic actuation, we now turn to dipole--dipole interactions. 
Helical rods with intra-rod magnetic attraction have been experimentally realized by \citet{sano2022kirchhoff}, and later numerically studied by \citet{sano2022reduced} and \citet{huang2023discrete}. 
\citet{sano2022reduced} added a dipole--dipole potential energy term to the Kirchhoff-like theory developed by \citet{sano2022kirchhoff}. 
Meanwhile, \citet{huang2023discrete} developed a model based on discrete differential geometry (DDG). 
In both these approaches, the continuous rod is discretized into a set of nodes and edge vectors. 
Unlike in the BPM, the nodes do not rotate; instead, the edge vectors rotate relative to some material frame.
Therefore, the dipoles are assigned to the edge vectors. 

In \citet{sano2022kirchhoff}, the hMRE was cast into a helical shape in its unstrained configuration and magnetized along the screw axis. 
If the dipole--dipole attraction is sufficiently strong, the helix will initially adopt a contracted configuration. 
Then applying a constant-gradient magnetic field along the screw axis produces an approximately constant body force on each discrete segment. 
By varying the strength of $\nabla \bm{B}$, the helix can be stretched or compressed along this axis. 
Notably, a hysteretic response over the loading--unloading cycle can be programmed by tuning the strength of the dipole--dipole interactions.
Hysteresis arises because the field gradient required to detach the helix exceeds the value at which it contracts.

A helix of radius $R$ and pitch $\psi$ can be arc-length-parametrized by $\left\langle R \cos{(Ks)}, R \sin{(Ks)}, s \cos{\psi} \right\rangle$, where $K = R^{-1} \sin{\psi}$. 
Here, $K = \sqrt{\kappa^2 + \tau^2}$, where $\kappa = K\sin{\psi}$ is the curvature and $\tau = K\cos{\psi}$ is the torsion. 
We consider the example in Fig.~2 of \citet{sano2022reduced} with the following helix geometry: $L = 103~\unit{mm}$, $L/(2\pi R) = 1.64$, and $\tilde{\psi} := \psi/(\pi/2) =0.96$. 
The helix was discretized into $N = 103$ bonds, making the bond length approximately $l = 1~\unit{mm}$. 
The rod diameter is $d = 2~\unit{mm}$. 
Unlike the previous example, where adding more particles generally improves accuracy, with dipole--dipole interactions the particles ideally should represent cylindrical segments of aspect ratio $d/l = \sqrt{4/3}$~\citep{petruska2012optimal}. 
However, since we use a granular contact model, more particles also lead to a smoother contact surface. 

\begin{figure}[!htpb]
    \centering
    \includegraphics[width=\linewidth]{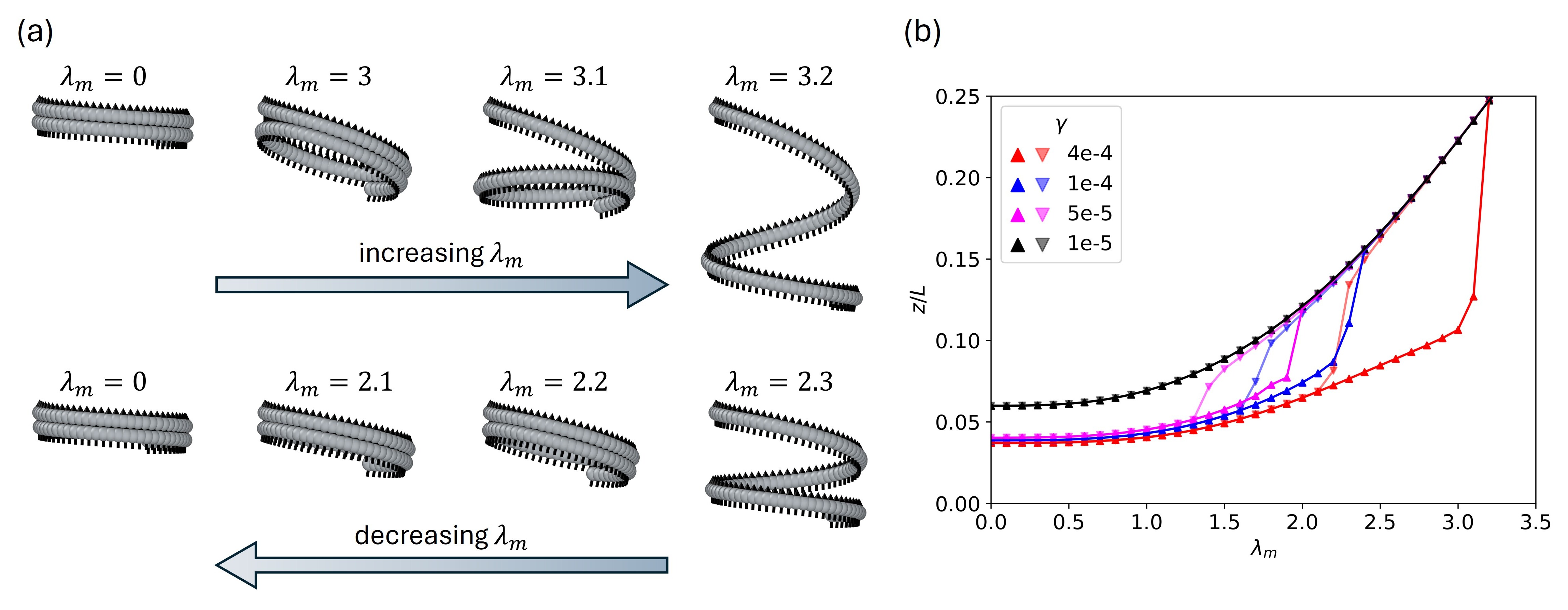}
    \caption{Magnetic helical rods with dipole--dipole interactions. (a)~Simulation snapshots for the helix with $\gamma = 4 \times 10^{-4}$. The top row shows notable configurations for increasing $\lambda_m$, and the bottom row for decreasing $\lambda_m$. (b)~The normalized distance as a function of $\lambda_m$ (external field strength) for different values of $\gamma$ (dipole--dipole interaction strength).}
    \label{fig:magnetic-helix}
\end{figure}

The particle at the base of the helix was clamped (topmost particle in Fig.~\ref{fig:magnetic-helix}), while the remaining particles evolved freely. 
A magnetic scalar potential of the form $\psi = -(b/\mu_0) \bigl[(x^2 + y^2)/4 - z^2/2\bigr]$ was applied, where $b$ is a variable prefactor, the $z$-axis is aligned with the screw axis, $z = 0$ at the clamped base, and the origin lies on the helix centerline. 
\citet{sano2022reduced} defined two dimensionless parameters that measure the magnetic forces relative to the elastic force: $\lambda_m = L [M b A / (E I)]^{1/3}$ for the external actuation and $\gamma = \mu_0 M^2 / E $ for the dipole--dipole interactions. 
Therefore, the prefactor can be written as 
\begin{equation*}
    b = \sqrt{\frac{\mu_0}{\gamma E}} \frac{E I}{A} \left(\frac{\lambda_m}{L}\right)^3 .
\end{equation*}
The simulation proceeded by quasi-statically ramping the dimensionless magnetization parameter $\lambda_m$ from $0$ to $3.5$ in increments of $0.1$. 
After reaching the maximum field strength, $\lambda_m$ was decreased back to zero following the same protocol. 
For the helix with the strongest dipole--dipole interactions ($\gamma = 4 \times 10^{-4}$), simulation snapshots at various $\lambda_m$ values are shown in Fig.~\ref{fig:magnetic-helix}(a). 
The normalized distance as a function of $\lambda_m$ for different $\gamma$ values is shown in Fig.~\ref{fig:magnetic-helix}(b). 

Our simulation results are consistent with Sano's~\citep{sano2022reduced}. 
The model captures the hysteresis and the correct critical values $\lambda_m^+$ and $\lambda_m^-$ for the detachment and contraction. 
However, we did not obtain a good quantitative match. 
Some of the discrepancy may be attributed to a difference in setup. 
For example, the initial displacement for the contracted state of the magnetized helices is around half of the value reported by Sano. 
We also observed rotation before the helix detaches, which can be seen in the $\lambda_m = 3$ snapshot in Fig.~\ref{fig:magnetic-helix}(a). 
\citet{huang2023discrete} also reported quantitative differences in their simulations using a DDG-based rod model and discussed possible causes of these discrepancies.

\subsection*{Fluid--structure interaction}

\begin{figure}[!htpb]
    \centering
    \includegraphics[width=0.9\linewidth]{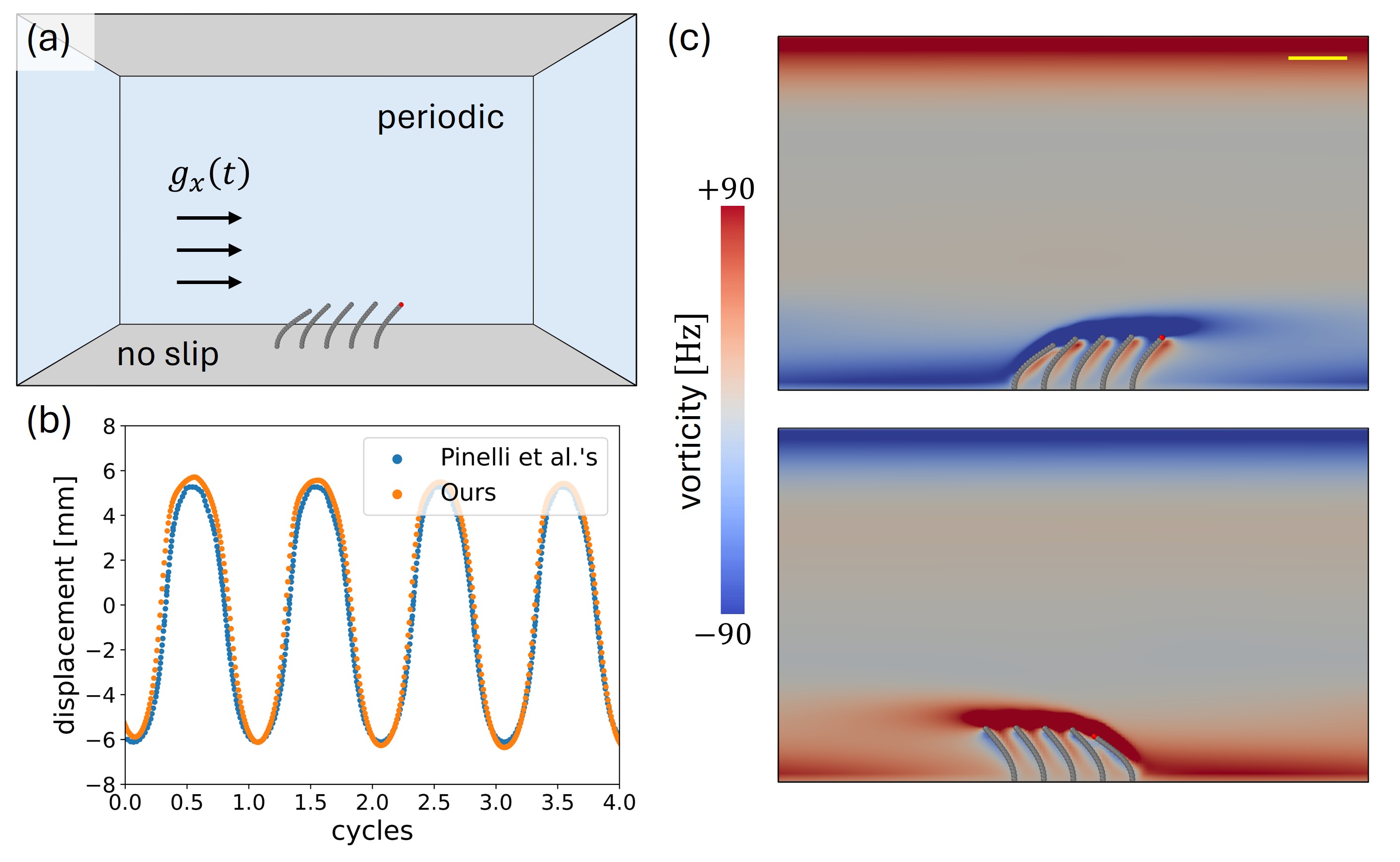}
    \caption{Filaments in an oscillatory flow. (a)~Simulation setup. (b)~Displacement of the rightmost filament's tip (highlighted in red in (a)) over successive cycles, compared with the results of \citet{pinelli2017pelskin}. (c)~The out-of-plane component of the vorticity in the plane containing the filaments during the maximum and minimum displacement. The scale bar (yellow) is $L = 10~\unit{mm}$.}
    \label{fig:filaments-oscillating-flow}
\end{figure}

To demonstrate multiphysics capability, we couple the BPM to a lattice Boltzmann fluid solver via the immersed boundary method (\textbf{Methods}).
To validate the coupling, we simulate five elastic filaments in an oscillatory flow, a setup studied experimentally and numerically by \citet{pinelli2017pelskin} (Fig.~\ref{fig:filaments-oscillating-flow}(a)). 
Each filament is a cylinder with diameter $d = 1~\unit{mm}$ and length $L = 10~\unit{mm}$, discretized into $N = 16$ bonds.
The filaments have Young's modulus $E = 1.23~\unit{MPa}$, Poisson's ratio $\nu_s = 0.3$, and density $\rho_s = 0.97~\unit{g}/\unit{cm}^3$.
They are clamped to the lower wall ($y = 0$) of a $10L \times 6L \times 5L$ channel at a spacing of $L/2$ along $x$, where the middle filament is at the center.
The fluid has kinematic viscosity $\nu_f = 0.1~\unit{mm}^2/\unit{ms}$ and density $\rho_f = 1.26~\unit{g}/\unit{cm}^3$, yielding a Reynolds number $\mathrm{Re} = u_{\max}L/\nu_f = 60$ based on a peak velocity of $u_{\max} = 0.6~\unit{m}/\unit{s}$.
A sinusoidal body-force acceleration $g_x(t) = g_0 \sin(2 \pi f\,t)$ drives the flow, where $f = 1~\unit{Hz}$ and $g_0 = u_{\max} \cdot 2\pi f = 3.8 \times 10^{-3}~\unit{mm}/\unit{ms}^2$ from the oscillatory Stokes flow solution.
The LBM grid has spacing $\Delta x = 0.625~\unit{mm}$ ($160 \times 96 \times 80$ nodes), which is the same as the BPM bond length.
The stiff BPM springs require sub-cycling at $N_\text{sub}=21$ DEM substeps per LBM step.
No-slip walls bound the channel in $y$; the remaining directions are periodic.
Fig.~\ref{fig:filaments-oscillating-flow}(b) shows the horizontal tip displacement of the rightmost filament over successive cycles, excluding the initial two cycles during which the filaments have not yet reached steady oscillation.
Our results agree well with \citet{pinelli2017pelskin} in both amplitude and frequency.
The horizontal axis is shifted to align the phase with the reference data, as our body-force-driven flow begins at a displacement minimum at the start of each cycle.
Fig.~\ref{fig:filaments-oscillating-flow}(c) shows the out-of-plane vorticity in the plane of the filaments at maximum and minimum tip displacement.

\subsection*{Fluid pumping by magnetic cilia arrays}\label{sec:magnetic-cilia-array}

\begin{figure}[!htpb]
    \centering
    \includegraphics[width=\linewidth]{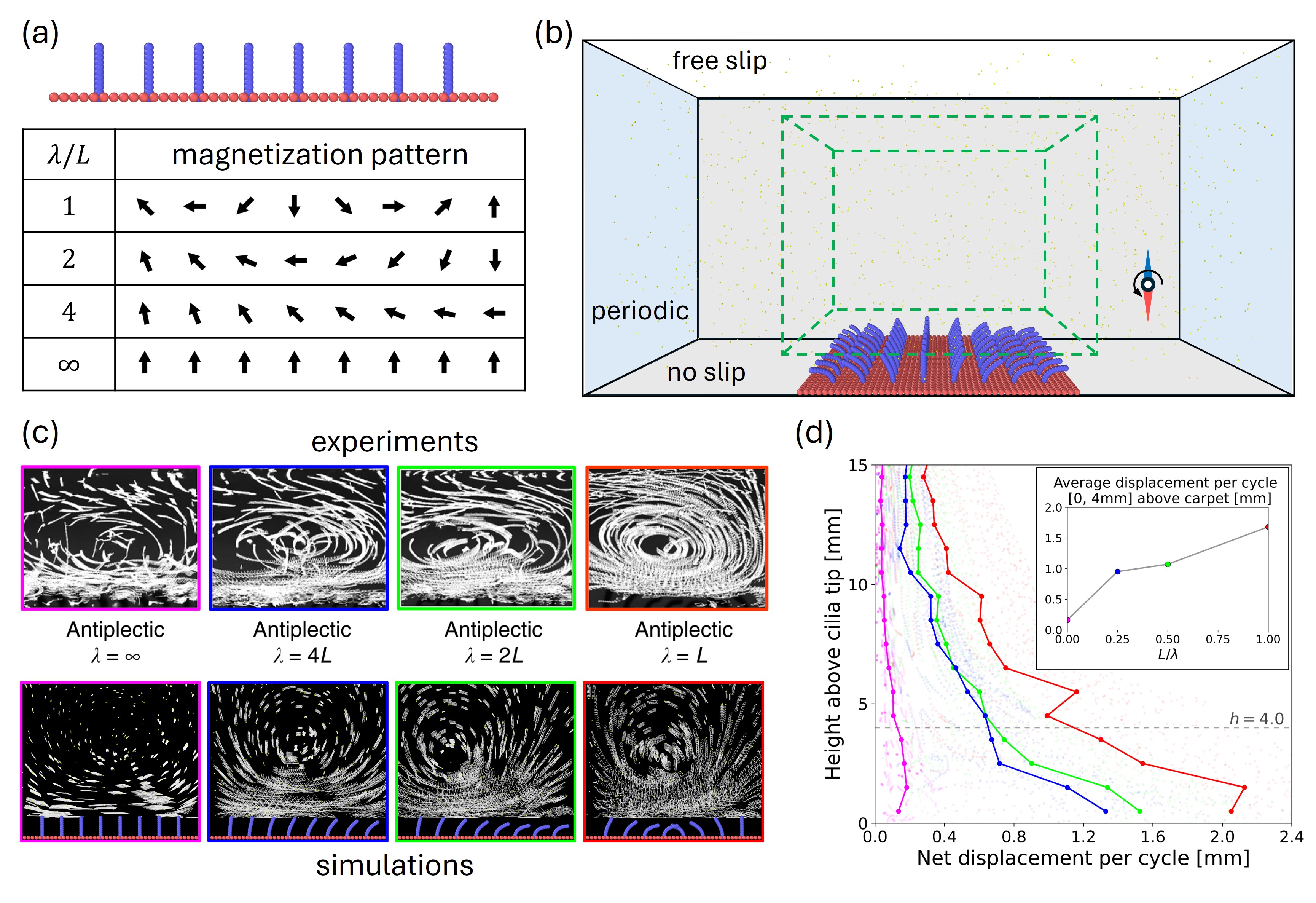}
    \caption{Fluid pumping by magnetic cilia arrays. (a)~Side view of the cilia array and representative magnetization patterns, controlled by the wavelength $\lambda$. Each row of cilia (same $y$, into the page) has the same magnetization, whose direction is indicated by the corresponding arrow in the table. The plate (red particles) lies in the $xy$-plane and is not included in the simulation, serving only for visualization. (b)~The cilia array is immersed in an open channel. A rotating uniform magnetic field is applied to actuate the cilia, and tracer particles are added to visualize the flow pattern. (c)~Comparison of tracer trajectories between experimental images of \citet{Gu2020} and our simulations. (d)~The average net displacement per cycle at various heights above the cilia tip. The colors correspond to those in (c). The inset shows the average displacement for the region $[0, 4~\unit{mm}]$ above the cilia tip.}
    \label{fig:magnetic-cilia-array}
\end{figure}

Using the developed LBM--IBM framework, we simulate fluid transport driven by a magnetic cilia array, an example from \citet{Gu2020} that has since been the subject of several numerical studies \citep{jiang2023numerical,huang2023modeling,tekinalp2025self}.
An $N_{cx} \times N_{cy} = 8 \times 8$ array of flexible cilia of length $L_c = 4~\unit{mm}$, spaced $D_\text{sep} = L_c$ apart, is placed in a large container. 
Each cilium has a circular cross-section of diameter $d = 0.8~\unit{mm}$ and extends in the $z$ direction. 
The material properties are: Young's modulus $E = 185~\unit{kPa}$, shear modulus $G = 61.6~\unit{kPa}$, and density $\rho_s = 2.39~\unit{g}/\unit{cm}^3$.
Each cilium's magnetization depends on its $x$ position, $\bm{M} = M (\cos\theta \, \hat{\mathbf{x}} + \sin\theta \, \hat{\mathbf{z}})$, where $M = 20~\unit{kA}/\unit{m}$ and $\theta(x) = 2 \pi x / \lambda + \theta_0$ for an antiplectic (wave direction opposite to the cilia power stroke direction) metachronal wave pattern. 
Here, we study the pumping behavior as a function of the wavelength $\lambda$, which \citet{Gu2020} showed affects the pumping efficiency. 
Fig.~\ref{fig:magnetic-cilia-array}(a) shows four different magnetization patterns corresponding to $\lambda/L = 1, \, 2, \, 4, \,\text{and}\, \infty$ (and $\theta_0 = 90^\circ$), where $L := N_{cx} D_\text{sep} = 32~\unit{mm}$. 
The structure is immersed in 99\% glycerol ($\nu_f = 0.9~\unit{mm}^2/\unit{ms}$, $\rho_f = 1.26~\unit{g}/\unit{cm}^3$).

In the simulations, each cilium was discretized into $N = 10$ bonds with the bottom particle clamped. 
Dipole--dipole interactions between cilia were neglected in this study; according to \citet{Gu2020}, these become significant when the cilia are densely packed.
A $90 \times 36 \times 45$ LBM grid with spacing $\Delta x = 1~\unit{mm}$ was used. 
Periodic boundary conditions were applied along $x$, no-slip along $y$ and the bottom $z$ walls, and free-slip on the top $z$ wall to approximate their setup.
A schematic of the simulation setup is shown in Fig.~\ref{fig:magnetic-cilia-array}(b).
To replicate their experimental procedure, $1000$ tracer particles (with diameter $200~\unit{\mu m}$ and density $1~\unit{g}/\unit{cm}^3$) were randomly seeded in the fluid domain and the net displacement per actuation cycle was recorded for particles inside the tracking region ($40 \times 36 \times 30$, $5~\unit{mm}$ above the structure, shown as the green box in Fig.~\ref{fig:magnetic-cilia-array}(b)).
A uniform magnetic field of magnitude $B = 80~\unit{mT}$, rotating in the $xz$-plane at a rate of $\omega = 30^\circ~\unit{s}^{-1}$, was applied. 
We simulated 10 actuation cycles and discarded the first cycle as a transient, as it was the only cycle whose displacement deviated appreciably from the rest.
The simulation for $\lambda = L$ is shown in Supplementary Video~3.

The simulated tracer trajectories are compared to the experimental images of \citet{Gu2020} in Fig.~\ref{fig:magnetic-cilia-array}(c).
The metachronal wave patterns generate recirculating flow above the array, qualitatively consistent with their experiments, though the detailed flow topology differs.
For each cycle, the net displacement of each tracer in the tracking region was recorded (faint dots in Fig.~\ref{fig:magnetic-cilia-array}(d)). 
The average displacement at each height was computed by binning the results into intervals of width matching the LBM grid spacing.
The inset shows the mean displacement at heights below $4~\unit{mm}$.
Consistent with their observation, the net transport decreases with increasing wavelength $\lambda$; however, the difference between $\lambda = 2L$ and $\lambda = 4L$ is not significant given the scatter in the data.
Moreover, this scalar measure does not account for the flow direction or the presence of the eddies, and should therefore be interpreted only as a qualitative indicator of pumping performance.

\section*{Discussion}\label{sec:disc}

We have developed a bonded-particle model that unifies large deformations, contact, and long-range magnetic interactions for slender rods within a single discrete-element framework.
Across all examples---extreme twisting, magnetized beam deflection, dipole-driven mechanical hysteresis, and fluid--structure interaction---the model shows good agreement with experimental, analytical, and numerical references.
However, the BPM comes with trade-offs.
The first concerns discretization, which depends on the system considered.
For problems governed only by external magnetic loading, the BPM can be accurate with as few as $10$ bonds, as is the case with a cantilever beam under an end load \citep{chen_comparative_2022,guo2013validation,nguyen2013validation}; we verify this with a convergence analysis in Supplementary Information~\ref{S-sec:static-beam-convergence}.
Here, the discretization need only be fine enough to capture the local distribution of magnetic force and torque along the rod.
For instance, the magnetized beam in Fig.~\ref{fig:magnetic-beams}(a) reduces to a cantilever beam problem under a slope-dependent moment and body force, and a convergence study in Supplementary Information~\ref{S-sec:magnetic-beam-convergence} confirms that the tip deflection converges linearly with increasing number of bonds.
This is lower than the quadratic or higher-order convergence typical of finite element and geometrically exact rod discretizations, and it represents the principal numerical trade-off of the BPM approach.
The advantage, however, is that the particle-level description makes it straightforward to retrofit additional multiphysics interactions---magnetic, fluidic, granular, or otherwise---without modifying the structural formulation.
We note that the bond length is the only free parameter in the spring constants; the material and geometric properties are determined by the system.

For the magnetic helical rods, the discretization is no longer unconstrained: the point-dipole approximation becomes inaccurate when individual segments are very long or very flat.
Therefore, the bond length is constrained by the rod's physical diameter. 
Similarly, for the granular contact model used in the extreme twisting examples, the particle diameter is set to the rod diameter to represent the physical contact surface, which in turn requires the bond length to be smaller than the particle diameter to prevent other particles from passing through the gaps between bonded particles. 
A bond-based contact model could be used to decouple the elastic and contact discretizations; for example, \citet{crassous2023DEM} combined a discrete elastic rod model with a Cundall--Strack contact model between cylindrical bonds.
These constraints on bond, magnetic, and contact resolution can conflict, and in general the different physical interactions may require separate discretizations.
For fluid--structure interaction, the immersed boundary coupling further requires the bond length and LBM grid spacing to be of comparable scale.
There is no universal choice of discretization; the optimal parameters depend on the dominant physics.
We refer the reader to the project page for specific guidance on each example.

The second concerns the dynamics.
All examples presented here employ dissipative dynamics: the quasi-static examples use artificial viscous damping to reach equilibrium, while the fluid-coupled examples are physically damped by the surrounding fluid.
Strict energy conservation was therefore neither required nor enforced.
Applying the BPM to conservative Hamiltonian dynamics would require addressing sources of energy drift. 
Most importantly, the energy in Eq.~\eqref{eq:total-energy} is only an approximation, and the rotation decomposition does not straightforwardly yield a global energy from which forces and torques can be obtained as gradients with respect to particle positions and orientations; addressing this is a key direction for future work.
Moreover, the time-integration treats each particle as a solid sphere, which has a different moment of inertia from the cylindrical segment it represents.
For quasi-static problems, this discrepancy is inconsequential, as inertial terms vanish at equilibrium and the static configuration is determined entirely by the bond forces and torques, which are independent of particle shape.
For dynamic problems, this provides an additional constraint on the discretization (Supplementary Information~\ref{S-sec:dynamic-beam-convergence}).
This limitation could be lifted by allowing user-defined diagonal inertia tensors.
Alternatively, cylindrical (super-ellipsoidal) particles can be used in LAMMPS to provide a more accurate moment of inertia and contact model.

Although the examples in this work are restricted to rod-like structures, the bonded-particle framework is not.
The BPM was originally developed for three-dimensional bond networks in granular and cementitious materials, and the co-rotational beam bonds introduced here can connect particles in arbitrary topologies---such as those in architected and woven metamaterials \citep{DESHPANDE20011747,wang2021structured,yang2025_lattice,zhou2025PAMs,carton2026woven}.
In the future, extending the formulation to shell-like and solid structures will allow us to simulate more complex systems.

\section*{Methods}\label{sec:BPM}

\subsection*{Magnetic interactions}

Under an external magnetic field $\bm{B}$, a dipole $\bm{m}$ has energy $U^\text{ext} = -\bm{m} \cdot \bm{B}$, from which the force $\bm{f} = \nabla(\bm{m} \cdot \bm{B})$ and torque $\bm{\tau} = \bm{m} \times \bm{B}$ follow. 
The pairwise interaction energy between two dipoles $\bm{m}_i$ and $\bm{m}_j$ separated by $\bm{r}_{ij}$ is
\begin{equation}
U^\text{dip}_{ij} = \frac{\mu_0}{4\pi r_{ij}^3} \left[ \bm{m}_i \cdot \bm{m}_j - 3 (\bm{m}_i \cdot \hat{\bm{r}}_{ij})(\bm{m}_j \cdot \hat{\bm{r}}_{ij}) \right],
\end{equation}
from which the forces and torques follow by differentiation.
While we focus on ferromagnetic rods, the BPM readily accommodates other magnetization models, including paramagnetic and superparamagnetic models common in soft matter and microscale systems \citep{dreyfus2005microscopic,cebers2005flexible,gerbal2015magnetoelastic,erb_2016_actuating}.

\subsection*{Damping}

Our bond model supports optional damping for each deformation mode. 
Radial damping is applied using the relative velocity projected along the bond axis, $f_r \mathrel{+}= g_r \, (\Delta \bm{v} \cdot \hat{\bm{r}}_f)$. 
For the angular deformation modes, strain rates are estimated by finite differences of the stored deformation measures: $\dot{\gamma} = (\gamma^{n} - \gamma^{n-1})/\Delta t$ for the shear angle, and similarly for the twist $\dot{\psi}$ and bending $\dot{\theta}$ angles. 
Rate-proportional terms are then added to the corresponding elastic forces and torques, $f_s \mathrel{+}= g_s \, r_f \, \dot{\gamma}$ (with the associated $\tau_s$ updated accordingly), $\tau_t \mathrel{+}= g_t \, \dot{\psi}$, and $\tau_b \mathrel{+}= g_b \, \dot{\theta}$, where the $g$ coefficients are the corresponding damping factors.
The bond-level damping targets high-frequency oscillations between neighboring particles, while global viscous damping dissipates low-frequency kinetic energy and aids convergence toward quasi-static equilibrium. 

\subsection*{Contact}

Since the BPM originated in geomechanics, our model benefits from the readily available, well-established granular contact models. 
Our formulation is therefore applicable to elasto-granular coupling problems~\citep{schunter2018elastogranular,manning2023essay,pol2025granular}, which we leave for future work.  
In the twisted rod examples, granular interactions prevent rod self-intersection, while in the magnetic helix example they enforce a minimum separation to avoid the singularity in the dipole--dipole potential. 
Both cases use a Hertz--Mindlin contact model between non-bonded particles.

\subsection*{Lattice Boltzmann method}\label{sec:LBM}

The LBM recovers macroscopic flow quantities from the evolution of particle distribution functions via collision and streaming steps on a fixed Cartesian grid.
Owing to its locality and flexibility, LBM exhibits good parallel scalability and is naturally suited for coupling with our particle-based rod model and the domain-decomposition framework used in LAMMPS. 
We develop a custom lightweight LBM--IBM package in LAMMPS with assistance from an AI language model (Claude, Anthropic). All generated code was reviewed, tested, and validated by the authors.
The primary goal is to demonstrate the LBM--BPM coupling.
More complex applications may benefit from established implementations available within LAMMPS \citep{tan2018parallel,ye2020openfsi,denniston2022lammps}.

A detailed treatment of the LBM can be found in~\citet{kruger2017lattice}; here we summarize the essential features relevant to the present work. 
The LBM solves the discrete Boltzmann equation by evolving particle distribution functions $f_i(\bm{x}, t)$ on a regular lattice, where each $f_i$ represents the probability of finding particles moving with discrete velocity $\bm{e}_i$. 
The evolution proceeds in two steps: collision, in which distributions at each node relax toward a local Maxwell--Boltzmann equilibrium $f_i^{\mathrm{eq}}$ under the Bhatnagar--Gross--Krook (BGK) approximation \citep{bhatnagar1954model}; and streaming, in which post-collision distributions propagate to neighboring nodes along their respective velocity directions. 
The combined update reads
\begin{equation}
  f_i(\bm{x} + \bm{e}_i \Delta t,\, t + \Delta t)
  = f_i(\bm{x}, t)
  - \frac{1}{\tau}\bigl[f_i(\bm{x}, t) - f_i^{\mathrm{eq}}(\bm{x}, t)\bigr]
  + S_i,
\end{equation}
where $\tau$ is the relaxation time related to the kinematic viscosity by $\nu = c_s^2(\tau - \tfrac{1}{2})\Delta t$, $c_s = 1/\sqrt{3}$ is the lattice speed of sound, and $S_i$ is the Guo forcing term~\citep{guo2002discrete}. 
External body forces are incorporated through the Guo scheme, which adds a source term to the collision step,
\begin{equation}
  S_i = \left(1 - \frac{1}{2\tau}\right) w_i
  \left[
    \frac{\bm{e}_i \cdot \bm{F}}{c_s^2}
    + \frac{(\bm{e}_i \cdot \bm{u})(\bm{e}_i \cdot \bm{F})}{c_s^4}
    - \frac{\bm{u} \cdot \bm{F}}{c_s^2}
  \right],
\end{equation}
and defines the macroscopic velocity at the half-timestep to ensure second-order accuracy in the recovered Navier--Stokes equations. 
Macroscopic density and momentum are obtained as
\begin{equation}
  \rho = \sum_i f_i, \qquad
  \rho\bm{u} = \sum_i f_i\,\bm{e}_i + \tfrac{1}{2}\bm{F}\,\Delta t.
\end{equation}
In this work we employ the three-dimensional D3Q19 lattice, where in each
step, the distribution functions stream along 19 possible directions (one rest, six
face-connected, and twelve edge-connected neighbors). 

\subsection*{Immersed boundary coupling}

In the IBM~\citep{peskin2002immersed}, Lagrangian DEM particles exchange momentum with the Eulerian fluid lattice through a regularized delta function. 
At each coupling step, the fluid velocity at a particle position $\bm{X}$ is obtained by interpolation,
\begin{equation}
  \bm{U}(\bm{X}) = \sum_{\bm{x}} \bm{u}(\bm{x})\,
  \delta_h(\bm{x} - \bm{X})\, \Delta x^D,
\end{equation}
where $\delta_h$ is a smoothed discrete delta function, $\Delta x$ is the lattice spacing, and $D$ is the spatial dimension. 
A penalty force is then computed to drive the particle velocity toward the local fluid velocity,
\begin{equation}
  \bm{F}_p = \kappa (\bm{U} - \bm{v}_p),
\end{equation}
where $\kappa$ is the penalty stiffness. 
The reaction force $-\bm{F}_p$ is spread back onto the fluid lattice via the same delta function,
\begin{equation}
  \bm{f}(\bm{x}) = -\sum_p \bm{F}_p\,
  \delta_h(\bm{x} - \bm{X}_p)\, \frac{\Delta V_p}{\Delta x^D},
\end{equation}
where $\Delta V_p$ is the Lagrangian volume element associated with the particle. 

\begin{figure}[!htpb]
    \centering
    \includegraphics[width=0.8\linewidth]{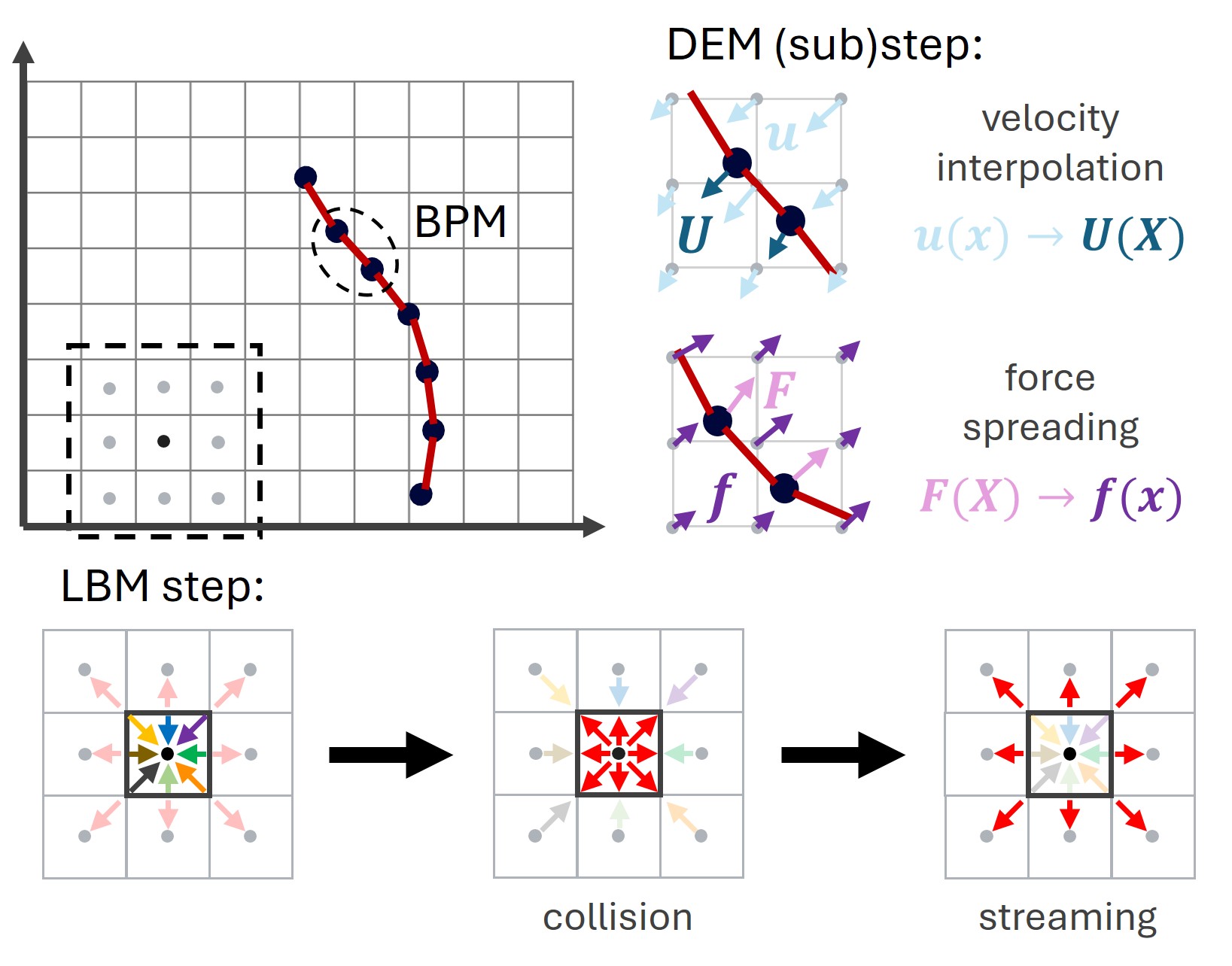}
    \caption{Schematic of the LBM--BPM coupling. A rod is immersed in a fluid grid. Each LBM step updates the macroscopic fluid properties by collision and streaming of the distribution functions. For each DEM (sub)step, the fluid velocity is interpolated to the current position of the particle. A penalty force is computed from the difference between the interpolated fluid velocity and the actual particle velocity, where equal and opposite forces are applied to the particle and spread into the fluid grid.}
    \label{fig:LBM-BPM}
\end{figure}

Typical BPM applications require time steps smaller than the fluid time step ($\Delta t_{\text{DEM}} < \Delta t_{\text{LBM}}$) for stability, such that multiple DEM steps are performed per LBM step. 
Following~\citet{tretyakov2017improved,vlogman2024efficient}, the fluid velocity field is interpolated once at the beginning of each LBM cycle, but the penalty force---with stiffness $\kappa = m_p / \Delta t_{\text{LBM}}$, where $m_p$ is the particle mass---is recomputed at every DEM substep using the particle's current velocity. 
The reaction force spread onto the fluid lattice is scaled by $1/N_\text{sub}$, where $N_\text{sub} = \Delta t_{\text{LBM}} / \Delta t_{\text{DEM}}$ is the number of substeps, so that momentum is conserved over each LBM cycle. 
We do not update the velocity field during the DEM substep; this option could be added in the future.
The present implementation treats solid nodes as point particles, neither transmitting nor receiving torques from the fluid.
A schematic of the LBM--BPM coupling is shown in Fig.~\ref{fig:LBM-BPM}, using a D2Q9 lattice for illustration.

\backmatter

\section*{Acknowledgments}
GA and TZ gratefully acknowledge the National Science Foundation for 
financial support through Grant Nos.\ CMMI-1847149 and OIA-2428643. 
GA gratefully acknowledges the NSF INTERN supplemental funding 
program under Grant No.\ CMMI-1847149 for supporting an internship 
at Sandia National Laboratories. 
Simulations were performed using the Zest cluster at Syracuse 
University, the Solo cluster at Sandia National Laboratories, and 
Expanse at the San Diego Supercomputer Center (SDSC) through 
allocation MSS170004 from the ACCESS program, which is 
supported by U.S.\ National Science Foundation grants \#2138259, 
\#2138286, \#2138307, \#2137603, and \#2138296.

This work was performed, in part, at the Center for Integrated Nanotechnologies, an Office of Science User Facility operated for the U.S. Department of Energy (DOE) Office of Science.
Sandia National Laboratories is a multi-mission laboratory managed and operated by National Technology \& Engineering Solutions of Sandia, LLC (NTESS), a wholly owned subsidiary of Honeywell International Inc., for the U.S. Department of Energy's National Nuclear Security Administration (DOE/NNSA) under contract DE-NA0003525. This written work is authored by an employee of NTESS. The employee, not NTESS, owns the right, title and interest in and to the written work and is responsible for its contents. Any subjective views or opinions that might be expressed in the written work do not necessarily represent the views of the U.S. Government. The publisher acknowledges that the U.S. Government retains a non-exclusive, paid-up, irrevocable, world-wide license to publish or reproduce the published form of this written work or allow others to do so, for U.S. Government purposes. The DOE will provide public access to results of federally sponsored research in accordance with the DOE Public Access Plan.

\section*{Author Contributions}
G.A. performed the investigation, developed the software, conducted formal analysis, prepared the figures, and wrote the original draft. J.T.C. contributed to software development and methodology and provided computational resources. C.D.S. contributed to formal analysis. T.Z. conceptualized the study, supervised the project, and acquired funding. J.T.C., C.D.S., and T.Z. reviewed and edited the manuscript. All authors reviewed the manuscript.

\section*{Competing interests}
The authors declare no competing financial or non-financial interests.

\section*{Additional information}
\bmhead{Supplementary materials} The online version contains supplementary material available at \comment{(TBD)}.

\section*{Data availability}
The LAMMPS scripts required to reproduce the results presented in this study are available at \url{https://gsalkuin.github.io/lammps-bpm-rods/}. 
The scripts require the updated \texttt{bond\_style bpm/rotational}, which is available in LAMMPS as of the 4 July 2026 feature release (\url{https://docs.lammps.org/bond_bpm_rotational.html}).
The custom LBM--IBM package is available at \url{https://github.com/tengzhang48/CoupLB}.

\bigskip

\bibliography{sn-bibliography}

@article{peyer2013bio,
  title={Bio-inspired magnetic swimming microrobots for biomedical applications},
  author={Peyer, Kathrin E and Zhang, Li and Nelson, Bradley J},
  journal={Nanoscale},
  volume={5},
  number={4},
  pages={1259--1272},
  year={2013},
  publisher={Royal Society of Chemistry},
  doi={10.1039/c2nr32554c}
}

@article{yang2020magnetic,
  title={Magnetic actuation systems for miniature robots: A review},
  author={Yang, Zhengxin and Zhang, Li},
  journal={Advanced Intelligent Systems},
  volume={2},
  number={9},
  pages={2000082},
  year={2020},
  publisher={Wiley Online Library},
  doi={10.1002/aisy.202000082}
}

@article{hwang2020review,
  title={A review of magnetic actuation systems and magnetically actuated guidewire-and catheter-based microrobots for vascular interventions},
  author={Hwang, Junsun and Kim, Jin-young and Choi, Hongsoo}, 
  journal={Intelligent Service Robotics},
  volume={13},
  number={1},
  pages={1--14},
  year={2020},
  publisher={Springer},
  doi={10.1007/s11370-020-00311-0}
}

@article{dreyfus2024dexterous,
  title={Dexterous helical magnetic robot for improved endovascular access},
  author={Dreyfus, Roland and Boehler, Quentin and Lyttle, Sean and Gruber, Philipp and Lussi, Jonas and Chautems, Christophe and Gervasoni, Simone and Berberat, Jatta and Seibold, Dominic and Ochsenbein-K{\"o}lble, Nicole and others},
  journal={Science Robotics},
  volume={9},
  number={87},
  pages={eadh0298},
  year={2024},
  publisher={American Association for the Advancement of Science},
  doi={10.1126/scirobotics.adh0298}
}

@article{DESHPANDE20011747,
title = {Effective properties of the octet-truss lattice material},
journal = {Journal of the Mechanics and Physics of Solids},
volume = {49},
number = {8},
pages = {1747-1769},
year = {2001},
issn = {0022-5096},
doi = {10.1016/S0022-5096(01)00010-2},
author = {V.S. Deshpande and N.A. Fleck and M.F. Ashby}
}

@article{wang2021structured,
  title={Structured fabrics with tunable mechanical properties},
  author={Wang, Yifan and Li, Liuchi and Hofmann, Douglas and Andrade, Jos{\'e} E and Daraio, Chiara},
  journal={Nature},
  volume={596},
  number={7871},
  pages={238--243},
  year={2021},
  publisher={Nature Publishing Group UK London},
  doi={10.1038/s41586-021-03698-7}
}

@article{Aghamiri2025,
author = {Aghamiri, Seyyedmohammad and Sedaghati, Ramin},
title = {Magnetoactive Metamaterials: A State-of-the-Art Review},
journal = {Advanced Engineering Materials},
volume = {27},
number = {23},
pages = {e202501312},
doi = {https://doi.org/10.1002/adem.202501312},
year = {2025}
}

@article{zhou2025PAMs,
author = {Wenjie Zhou  and Sujeeka Nadarajah  and Liuchi Li  and Anna Guell Izard  and Hujie Yan  and Aashutosh K. Prachet  and Payal Patel  and Xiaoxing Xia  and Chiara Daraio },
title = {{3D} polycatenated architected materials},
journal = {Science},
volume = {387},
number = {6731},
pages = {269-277},
year = {2025},
doi = {10.1126/science.adr9713}
}

@article{carton2026woven,
  title     = "Design framework for programmable three-dimensional woven
               metamaterials",
  author    = "Carton, Molly and Surjadi, James Utama and Aymon, Bastien F G
               and Xu, Ling and Portela, Carlos M",
  journal   = "Nature Communications",
  publisher = "Springer Science and Business Media LLC",
  volume    =  17,
  number    =  1,
  pages     = "1581",
  month     =  jan,
  year      =  2026,
  doi       = {10.1038/s41467-026-68298-3}
}

@article{yang2023_networks,
author = {Yang, Xinyan and Leng, Junqing and Sun, Cheng and Keten, Sinan},
title = {Self-Assembled Robust {2D} Networks from Magneto-Elastic Bars},
journal = {Advanced Materials Technologies},
volume = {8},
number = {14},
pages = {2202189},
doi = {https://doi.org/10.1002/admt.202202189},
year = {2023}
}

@article{yang2025_lattice,
author = {Yang, Xinyan and Leng, Junqing and Sun, Cheng and Keten, Sinan},
title = {Highly Ordered {2D} Open Lattices Through Self-Assembly of Magnetic Units},
journal = {Advanced Functional Materials},
volume = {35},
number = {2},
pages = {2412326},
doi = {https://doi.org/10.1002/adfm.202412326},
year = {2025}
}

@article{gu2019quadrupole,
author = {Hongri Gu  and Quentin Boehler  and Daniel Ahmed  and Bradley J. Nelson },
title = {Magnetic quadrupole assemblies with arbitrary shapes and magnetizations},
journal = {Science Robotics},
volume = {4},
number = {35},
pages = {eaax8977},
year = {2019},
doi = {10.1126/scirobotics.aax8977},
}

@article{korpas2021temperature,
  title={Temperature-responsive multistable metamaterials},
  author={Korpas, Lucia M and Yin, Rui and Yasuda, Hiromi and Raney, Jordan R},
  journal={ACS Applied Materials \& Interfaces},
  volume={13},
  number={26},
  pages={31163--31170},
  year={2021},
  publisher={ACS Publications},
  doi={10.1021/acsami.1c07327}
}

@article{liang2022phase,
  title={Phase-transforming metamaterial with magnetic interactions},
  author={Liang, Xudong and Fu, Hongbo and Crosby, Alfred J},
  journal={Proceedings of the National Academy of Sciences},
  volume={119},
  number={1},
  pages={e2118161119},
  year={2022},
  publisher={National Academy of Sciences},
  doi={10.1073/pnas.2118161119}
}

@article{zou2023magneto,
  title={Magneto-thermomechanically reprogrammable mechanical metamaterials},
  author={Zou, Bihui and Liang, Zihe and Zhong, Dijia and Cui, Zhiming and Xiao, Kai and Shao, Shuang and Ju, Jaehyung},
  journal={Advanced Materials},
  volume={35},
  number={8},
  pages={2207349},
  year={2023},
  publisher={Wiley Online Library},
  doi={10.1002/adma.202207349}
}

@article{sun2026kirigami,
    author = {Sun, Haoze and Alkuino, Gabriel and Chi, Yinding and Zabila, Yevhen and Qing, Haitao and Makarov, Denys and Zhang, Teng and Yin, Jie},
    title = {Magnetic coupling transforms random snapping into ordered sequences in soft metamaterials},
    journal = {Science Advances},
    year = {2026},
    doi = {10.1126/sciadv.aec3182}
}

@article{park2023bioinspired,
  title={Bioinspired magnetic cilia: from materials to applications},
  author={Park, Seongjin and Choi, Geonjun and Kang, Minsu and Kim, Woochan and Kim, Jangho and Jeong, Hoon Eui},
  journal={Microsystems \& Nanoengineering},
  volume={9},
  number={1},
  pages={153},
  year={2023},
  publisher={Nature Publishing Group UK London},
  doi={10.1038/s41378-023-00611-2}
}

@article{erb_2016_actuating,
author = {Erb, Randall M. and Martin, Joshua J. and Soheilian, Rasam and Pan, Chunzhou and Barber, Jabulani R.},
title = {Actuating Soft Matter with Magnetic Torque},
journal = {Advanced Functional Materials},
volume = {26},
number = {22},
pages = {3859-3880},
doi = {10.1002/adfm.201504699},
year = {2016}
}

@article{lum_shape-programmable_2016,
author = {Guo Zhan Lum  and Zhou Ye  and Xiaoguang Dong  and Hamid Marvi  and Onder Erin  and Wenqi Hu  and Metin Sitti },
title = {Shape-programmable magnetic soft matter},
journal = {Proceedings of the National Academy of Sciences},
volume = {113},
number = {41},
pages = {E6007-E6015},
year = {2016},
doi = {10.1073/pnas.1608193113},
}

@article{Hu2018,
  author = {Hu, Wenqi and Lum, Guo Zhan and Mastrangeli, Massimo and Sitti, Metin},
  title = {Small-scale soft-bodied robot with multimodal locomotion},
  journal = {Nature},
  volume = {554},
  number = {7690},
  pages = {81--85},
  year = {2018},
  doi = {10.1038/nature25443}
}

@article{Gu2020,
  author = {Gu, Hongri and Boehler, Quentin and Cui, Haoyang and Secchi, Eleonora and Savorana, Giovanni and De Marco, Carmela and Gervasoni, Simone and Peyron, Quentin and Huang, Tian-Yun and Pane, Salvador and Hirt, Ann M. and Ahmed, Daniel and Nelson, Bradley J.},
  title = {Magnetic cilia carpets with programmable metachronal waves},
  journal = {Nature Communications},
  volume = {11},
  number = {1},
  pages = {2637},
  year = {2020},
  doi = {10.1038/s41467-020-16458-4}
}

@article{gu2023self,
  title={Self-folding soft-robotic chains with reconfigurable shapes and functionalities},
  author={Gu, Hongri and M{\"o}ckli, Marino and Ehmke, Claas and Kim, Minsoo and Wieland, Matthias and Moser, Simon and Bechinger, Clemens and Boehler, Quentin and Nelson, Bradley J},
  journal={Nature Communications},
  volume={14},
  number={1},
  pages={1263},
  year={2023},
  publisher={Nature Publishing Group UK London},
  doi={10.1038/s41467-023-36819-z}
}

@article{lee2023magnetically,
  title={Magnetically actuated fiber-based soft robots},
  author={Lee, Youngbin and Koehler, Florian and Dillon, Tom and Loke, Gabriel and Kim, Yoonho and Marion, Juliette and Antonini, Marc-Joseph and Garwood, Indie C and Sahasrabudhe, Atharva and Nagao, Keisuke and others},
  journal={Advanced Materials},
  volume={35},
  number={38},
  pages={2301916},
  year={2023},
  publisher={Wiley Online Library},
  doi={10.1002/adma.202301916}
}

@article{dreyfus2005microscopic,
  title={Microscopic artificial swimmers},
  author={Dreyfus, R{\'e}mi and Baudry, Jean and Roper, Marcus L and Fermigier, Marc and Stone, Howard A and Bibette, J{\'e}r{\^o}me},
  journal={Nature},
  volume={437},
  number={7060},
  pages={862--865},
  year={2005},
  publisher={Nature Publishing Group UK London},
  doi={10.1038/nature04090}
}

@article{wang_implementation_2006,
  title={Implementation of Particle-scale Rotation in the {3-D} Lattice Solid Model},
  author={Wang, Yucang and Abe, Steffen and Latham, Shane and Mora, Peter},
  journal={Pure and Applied Geophysics},
  volume={163},
  number={9},
  pages={1769--1785},
  year={2006},
  publisher={Springer},
  doi={10.1007/s00024-006-0096-0},
}

@article{wang2008macroscopic,
  title={Macroscopic elastic properties of regular lattices},
  author={Wang, Yucang and Mora, Peter},
  journal={Journal of the Mechanics and Physics of Solids},
  volume={56},
  number={12},
  pages={3459--3474},
  year={2008},
  publisher={Elsevier},
  doi={10.1016/j.jmps.2008.08.011},
}

@article{wang2009new,
  title={A new algorithm to model the dynamics of {3-D} bonded rigid bodies with rotations},
  author={Wang, Yucang},
  journal={Acta Geotechnica},
  volume={4},
  number={2},
  pages={117--127},
  year={2009},
  publisher={Springer},
  doi={10.1007/s11440-008-0072-1}
}

@article{wang_esys_particle_2009,
	author = {Wang, Yucang and Mora, Peter},
    title = {The {ESyS}\_{Particle}: {A} {New} 3-{D} {Discrete} {Element} {Model} with {Single} {Particle} {Rotation}. {In}: Advances in {Geocomputing}. {Lecture} {Notes} in {Earth} {Science}, vol 119. {Springer, Berlin, Heidelberg}.},
	year = {2009},
	doi = {10.1007/978-3-540-85879-9_6},
	pages = {183--228},
}

@article{potyondy2004bonded,
  title={A bonded-particle model for rock},
  author={Potyondy, David O and Cundall, Peter A},
  journal={International journal of rock mechanics and mining sciences},
  volume={41},
  number={8},
  pages={1329--1364},
  year={2004},
  publisher={Elsevier},
  doi={10.1016/j.ijrmms.2004.09.011}
}

@article{carmona2008fragmentation,
  title={Fragmentation processes in impact of spheres},
  author={Carmona, Humberto A and Wittel, Falk K and Kun, Ferenc and Herrmann, Hans J{\"u}rgen},
  journal={Physical Review E—Statistical, Nonlinear, and Soft Matter Physics},
  volume={77},
  number={5},
  pages={051302},
  year={2008},
  publisher={APS},
  doi={10.1103/PhysRevE.77.051302}
}

@article{andre2012discrete,
  title={Discrete element method to simulate continuous material by using the cohesive beam model},
  author={Andr{\'e}, Damien and Iordanoff, Ivan and Charles, Jean-luc and N{\'e}auport, J{\'e}r{\^o}me},
  journal={Computer methods in applied mechanics and engineering},
  volume={213},
  pages={113--125},
  year={2012},
  publisher={Elsevier},
  doi={10.1016/j.cma.2011.12.002}
}

@article{obermayr2013bonded,
  title={A bonded-particle model for cemented sand},
  author={Obermayr, Martin and Dressler, Klaus and Vrettos, Christos and Eberhard, Peter},
  journal={Computers and Geotechnics},
  volume={49},
  pages={299--313},
  year={2013},
  publisher={Elsevier},
  doi={10.1016/j.compgeo.2012.09.001}
}

@article{guo2013validation,
  title={Validation and time step determination of discrete element modeling of flexible fibers},
  author={Guo, Y and Wassgren, C and Hancock, B and Ketterhagen, W and Curtis, JJPT},
  journal={Powder technology},
  volume={249},
  pages={386--395},
  year={2013},
  publisher={Elsevier},
  doi={10.1016/j.powtec.2013.09.007}
}

@article{nguyen2013validation,
  title={Validation of partially flexible rod model based on discrete element method using beam deflection and vibration},
  author={Nguyen, Duong Hong and Kang, Namcheol and Park, Junyoung},
  journal={Powder Technology},
  volume={237},
  pages={147--152},
  year={2013},
  publisher={Elsevier},
  doi={10.1016/j.powtec.2013.01.038}
}

@article{lisjak2014review,
  title={A review of discrete modeling techniques for fracturing processes in discontinuous rock masses},
  author={Lisjak, A and Grasselli, G},
  journal={Journal of Rock Mechanics and Geotechnical Engineering},
  volume={6},
  number={4},
  pages={301--314},
  year={2014},
  publisher={Elsevier},
  doi={10.1016/j.jrmge.2013.12.007}
}

@article{potyondy2015bonded,
  title={The bonded-particle model as a tool for rock mechanics research and application: current trends and future directions},
  author={Potyondy, David Oskar},
  journal={Geosystem Engineering},
  volume={18},
  number={1},
  pages={1--28},
  year={2015},
  publisher={Taylor \& Francis},
  doi={10.1080/12269328.2014.998346}
}

@article{chen_comparative_2022,
	title = {A comparative assessment and unification of bond models in {DEM} simulations},
	volume = {24},
	issn = {1434-5021, 1434-7636},
	language = {en},
	number = {1},
	journal = {Granular Matter},
	author = {Chen, Xizhong and Peng, Di and Morrissey, John P. and Ooi, Jin Y.},
	year = {2022},
	pages = {29},
    doi={10.1007/s10035-021-01187-2}
}

@article{crassous2023DEM,
  title = {Discrete-element-method model for frictional fibers},
  author = {Crassous, J\'er\^ome},
  journal = {Phys. Rev. E},
  volume = {107},
  issue = {2},
  pages = {025003},
  numpages = {12},
  year = {2023},
  month = {Feb},
  publisher = {American Physical Society},
  doi = {10.1103/PhysRevE.107.025003}
}

@article{zhang2024rod,
  title={Rod-bonded discrete element method},
  author={Zhang, Kangrui and Yan, Han and Lu, Jia-Ming and Ren, Bo},
  journal={Graphical Models},
  volume={133},
  pages={101218},
  year={2024},
  publisher={Elsevier},
  doi={10.1016/j.gmod.2024.101218}
}

@article{clemmer2024soft,
  title={A soft departure from jamming: the compaction of deformable granular matter under high pressures},
  author={Clemmer, Joel T and Monti, Joseph M and Lechman, Jeremy B},
  journal={Soft Matter},
  volume={20},
  number={8},
  pages={1702--1718},
  year={2024},
  publisher={Royal Society of Chemistry},
  doi={10.1039/d3sm01373a}
}

@incollection{bergou2008discrete,
  title={Discrete elastic rods},
  author={Bergou, Mikl{\'o}s and Wardetzky, Max and Robinson, Stephen and Audoly, Basile and Grinspun, Eitan},
  booktitle={ACM SIGGRAPH 2008 papers},
  pages={1--12},
  year={2008},
  doi={10.1145/1399504.1360662}
}

@article{gazzola2018forward,
  title={Forward and inverse problems in the mechanics of soft filaments},
  author={Gazzola, M and Dudte, LH and McCormick, AG and Mahadevan, L},
  journal={Royal Society open science},
  volume={5},
  number={6},
  pages={171628},
  year={2018},
  publisher={The Royal Society Publishing},
  doi = {10.1098/rsos.171628}
}

@article{cebers2005flexible,
  title={Flexible magnetic filaments},
  author={C{\=e}bers, A},
  journal={Current opinion in colloid \& interface science},
  volume={10},
  number={3-4},
  pages={167--175},
  year={2005},
  publisher={Elsevier},
  doi={10.1016/j.cocis.2005.07.002}
}

@article{gerbal2015magnetoelastic,
author = {Fabien Gerbal  and Yuan Wang  and Florian Lyonnet  and Jean-Claude Bacri  and Thierry Hocquet  and Martin Devaud },
title = {A refined theory of magnetoelastic buckling matches experiments with ferromagnetic and superparamagnetic rods},
journal = {Proceedings of the National Academy of Sciences},
volume = {112},
number = {23},
pages = {7135-7140},
year = {2015},
doi = {10.1073/pnas.1422534112}
}

@article{wang2020hard,
  title={Hard-magnetic elastica},
  author={Wang, Liu and Kim, Yoonho and Guo, Chuan Fei and Zhao, Xuanhe},
  journal={Journal of the Mechanics and Physics of Solids},
  volume={142},
  pages={104045},
  year={2020},
  publisher={Elsevier},
  doi={10.1016/j.jmps.2020.104045}
}

@misc{yashraj_bhosale_2023_7658892,
  author       = {Yashraj Bhosale and
                  Arman Tekinalp},
  title        = {{MagnetoPyElastica}:  Open-source software for 
                   simulating magnetic {Cosserat} rods},
  month        = feb,
  year         = 2023,
  publisher    = {Zenodo},
  version      = {v.0.0.1.post1},
  doi          = {10.5281/zenodo.7658892}
}

@article{tekinalp2025self,
  title={Self-propelling, soft, and slender structures in fluids: {Cosserat} rods immersed in the velocity--vorticity formulation of the incompressible {Navier--Stokes} equations},
  author={Tekinalp, Arman and Bhosale, Yashraj and Cui, Songyuan and Chan, Fan Kiat and Gazzola, Mattia},
  journal={Computer Methods in Applied Mechanics and Engineering},
  volume={440},
  pages={117910},
  year={2025},
  publisher={Elsevier},
  doi={10.1016/j.cma.2025.117910}
}

@article{danas_stretch-independent_2024,
	title = {Stretch-independent magnetization in incompressible isotropic hard magnetorheological elastomers},
	volume = {191},
	issn = {0022-5096},
	journal = {Journal of the Mechanics and Physics of Solids},
	author = {Danas, Kostas and Reis, Pedro M.},
	month = oct,
	year = {2024},
	pages = {105764},
    doi={10.1016/j.jmps.2024.105764}
}

@article{lu2024mechanics,
  title={Mechanics of hard-magnetic soft materials: A review},
  author={Lu, Lu and Sim, Jay and Zhao, Ruike Renee},
  journal={Mechanics of Materials},
  volume={189},
  pages={104874},
  year={2024},
  publisher={Elsevier},
  doi={10.1016/j.mechmat.2023.104874}
}

@article{greaves2013poisson,
  title={{Poisson's} ratio over two centuries: challenging hypotheses},
  author={Greaves, G Neville},
  journal={Notes and Records of the Royal Society},
  volume={67},
  number={1},
  pages={37--58},
  year={2013},
  publisher={The Royal Society},
  doi={10.1098/rsnr.2012.0021}
}

@article{seung1988defects,
  title={Defects in flexible membranes with crystalline order},
  author={Seung, Hyunjune Sebastian and Nelson, David R},
  journal={Physical Review A},
  volume={38},
  number={2},
  pages={1005},
  year={1988},
  publisher={APS},
  doi={10.1103/physreva.38.1005}
}

@article{lloyd2007identification,
  title={Identification of spring parameters for deformable object simulation},
  author={Lloyd, Bryn and Sz{\'e}kely, G{\'a}bor and Harders, Matthias},
  journal={IEEE transactions on visualization and computer graphics},
  volume={13},
  number={5},
  pages={1081--1094},
  year={2007},
  publisher={IEEE},
  doi={10.1109/tvcg.2007.1055}
}

@article{golec2020hybrid,
  title={Hybrid 3D mass-spring system for simulation of isotropic materials with any Poisson’s ratio},
  author={Golec, Karolina and Palierne, J-F and Zara, Florence and Nicolle, St{\'e}phane and Damiand, Guillaume},
  journal={The Visual Computer},
  volume={36},
  number={4},
  pages={809--825},
  year={2020},
  publisher={Springer},
  doi={10.1007/s00371-019-01663-0}
}

@article{zhang2019deriving,
  title={Deriving a lattice model for neo-{Hookean} solids from finite element methods},
  author={Zhang, Teng},
  journal={Extreme Mechanics Letters},
  volume={26},
  pages={40--45},
  year={2019},
  publisher={Elsevier},
  doi={10.1016/j.eml.2018.11.007}
}

@article{ye2021magttice,
  title={{Magttice}: A lattice model for hard-magnetic soft materials},
  author={Ye, Huilin and Li, Ying and Zhang, Teng},
  journal={Soft Matter},
  volume={17},
  number={13},
  pages={3560--3568},
  year={2021},
  publisher={Royal Society of Chemistry},
  doi={10.1039/d0sm01662d}
}

@inproceedings{nealen2006physically,
  title={Physically based deformable models in computer graphics},
  author={Nealen, Andrew and M{\"u}ller, Matthias and Keiser, Richard and Boxerman, Eddy and Carlson, Mark},
  booktitle={Computer graphics forum},
  volume={25},
  number={4},
  pages={809--836},
  year={2006},
  organization={Wiley Online Library},
  doi={10.1111/j.1467-8659.2006.01000.x}
}

@article{simo1985finite,
  title={A finite strain beam formulation. The three-dimensional dynamic problem. Part I},
  author={Simo, Juan C},
  journal={Computer methods in applied mechanics and engineering},
  volume={49},
  number={1},
  pages={55--70},
  year={1985},
  publisher={Elsevier},
  doi={10.1016/0045-7825(85)90050-7}
}

@article{rankin1986element,
  author = {Rankin, C. C. and Brogan, F. A.},
  title = {An Element Independent Corotational Procedure for the Treatment of Large Rotations},
  journal = {Journal of Pressure Vessel Technology},
  volume = {108},
  number = {2},
  pages = {165-174},
  year = {1986},
  month = {05},
  issn = {0094-9930},
  doi = {10.1115/1.3264765}
}

@article{crisfield1990consistent,
  title={A consistent co-rotational formulation for non-linear, three-dimensional, beam-elements},
  author={Crisfield, Michael A},
  journal={Computer methods in applied mechanics and engineering},
  volume={81},
  number={2},
  pages={131--150},
  year={1990},
  publisher={Elsevier},
  doi={10.1016/0045-7825(90)90106-v}
}

@article{felippa2005unified,
  title={A unified formulation of small-strain corotational finite elements: {I}. Theory},
  author={Felippa, Carlos A and Haugen, Bjorn},
  journal={Computer Methods in Applied Mechanics and Engineering},
  volume={194},
  number={21-24},
  pages={2285--2335},
  year={2005},
  publisher={Elsevier},
  doi={10.1016/j.cma.2004.07.035}
}

@article{le2011efficient,
  title={Efficient formulation for dynamics of corotational {2D} beams},
  author={Le, Thanh-Nam and Battini, Jean-Marc and Hjiaj, Mohammed},
  journal={Computational Mechanics},
  volume={48},
  number={2},
  pages={153--161},
  year={2011},
  publisher={Springer},
  doi={10.1007/s00466-011-0585-6}
}

@article{le2014consistent,
  title={A consistent {3D} corotational beam element for nonlinear dynamic analysis of flexible structures},
  author={Le, Thanh-Nam and Battini, Jean-Marc and Hjiaj, Mohammed},
  journal={Computer Methods in Applied Mechanics and Engineering},
  volume={269},
  pages={538--565},
  year={2014},
  publisher={Elsevier},
  doi={10.1016/j.cma.2013.11.007}
}

@article{cottanceau2018finite,
  title={A finite element/quaternion/asymptotic numerical method for the {3D} simulation of flexible cables},
  author={Cottanceau, Emmanuel and Thomas, Olivier and V{\'e}ron, Philippe and Alochet, Marc and Deligny, Renaud},
  journal={Finite Elements in Analysis and Design},
  volume={139},
  pages={14--34},
  year={2018},
  publisher={Elsevier},
  doi={10.1016/j.finel.2017.10.002}
}

@article{grange2024co,
  title={Co-rotational {3D} beam element using quaternion algebra to account for large rotations: Formulation theory and static applications},
  author={Grange, St{\'e}phane and Bertrand, David},
  journal={International Journal of Solids and Structures},
  volume={293},
  pages={112746},
  year={2024},
  publisher={Elsevier},
  doi={10.1016/j.ijsolstr.2024.112746}
}

@article{yang20223d,
  title={A {3D} hard-magnetic rod model based on co-rotational formulations},
  author={Yang, Yifan and Li, Maoyuan and Xu, Fan},
  journal={Acta Mechanica Sinica},
  volume={38},
  number={9},
  pages={222085},
  year={2022},
  publisher={Springer},
  doi={10.1007/s10409-022-22085-x}
}

@book{reddy1993introduction,
  title={An introduction to the finite element method},
  author={Reddy, Junuthula Narasimha},
  year={1993},
  publisher={McGraw-Hill},
  address={Columbus},
  isbn={978-0070513556}
}

@book{hutton2003FEA,
  title={Fundamentals of Finite Element Analysis},
  author={Hutton, David},
  year={2003},
  publisher={McGraw-Hill},
  address={Columbus},
  isbn={978-0072922363}
}

@book{slaughter2012linearized,
  title={The linearized theory of elasticity},
  author={Slaughter, William S},
  year={2012},
  publisher={Birkh\"{a}user},
  address={Boston},
  doi={10.1007/978-1-4612-0093-2},
  isbn={978-1-4612-0093-2}
}

@book{dorfmann2014nonlinear,
  title={Nonlinear theory of electroelastic and magnetoelastic interactions},
  author={Dorfmann, Luis and Ogden, Ray W},
  volume={1},
  year={2014},
  publisher={Springer},
  address={New York},
  doi={10.1007/978-1-4614-9596-3},
  isbn={978-1-4614-9596-3}
}

@book{hibbeler2018mechanics,
  title={Mechanics of Materials},
  author={Hibbeler, R.C.},
  isbn={9781292178202},
  year={2018},
  publisher={Pearson},
  address={London}
}

@article{yan_comprehensive_2022,
	title = {A comprehensive framework for hard-magnetic beams: {Reduced}-order theory, {3D} simulations, and experiments},
	volume = {257},
	issn = {0020-7683},
	doi = {10.1016/j.ijsolstr.2021.111319},
	journal = {International Journal of Solids and Structures},
	author = {Yan, Dong and Abbasi, Arefeh and Reis, Pedro M.},
	year = {2022},
	pages = {111319}
}

@article{sano2022kirchhoff,
  title={A {Kirchhoff}-like theory for hard magnetic rods under geometrically nonlinear deformation in three dimensions},
  author={Sano, Tomohiko G and Pezzulla, Matteo and Reis, Pedro M},
  journal={Journal of the Mechanics and Physics of Solids},
  volume={160},
  pages={104739},
  year={2022},
  publisher={Elsevier},
  doi={10.1016/j.jmps.2021.104739}
}

@article{sano2022reduced,
  title={Reduced theory for hard magnetic rods with dipole--dipole interactions},
  author={Sano, Tomohiko G},
  journal={Journal of Physics A: Mathematical and Theoretical},
  volume={55},
  number={10},
  pages={104002},
  year={2022},
  publisher={IOP Publishing},
  doi={10.1088/1751-8121/ac4de2}
}

@article{huang2023discrete,
  title={A discrete model for the geometrically nonlinear mechanics of hard-magnetic slender structures},
  author={Huang, Weicheng and Liu, Mingchao and Hsia, K Jimmy},
  journal={Extreme Mechanics Letters},
  volume={59},
  pages={101977},
  year={2023},
  publisher={Elsevier},
  doi={10.1016/j.eml.2023.101977}
}

@article{huang2025tutorial,
  title={A tutorial on simulating nonlinear behaviors of flexible structures with the discrete differential geometry ({DDG}) method},
  author={Huang, Weicheng and Hao, Zhuonan and Li, Jiahao and Tong, Dezhong and Guo, Kexin and Zhang, Yingchao and Gao, Huajian and Hsia, K Jimmy and Liu, Mingchao},
  journal={Applied Mechanics Reviews},
  pages={1--88},
  year={2025},
  doi={10.1115/1.4069025}
}

@article{goss2005experiments,
  title={Experiments on snap buckling, hysteresis and loop formation in twisted rods},
  author={Goss, VGA and Van der Heijden, GHM and Thompson, JMT and Neukirch, S},
  journal={Experimental mechanics},
  volume={45},
  number={2},
  pages={101--111},
  year={2005},
  publisher={Springer},
  doi={10.1007/bf02428182}
}

@article{lazarus2013continuation,
  title={Continuation of equilibria and stability of slender elastic rods using an asymptotic numerical method},
  author={Lazarus, Arnaud and Miller, JT and Reis, Pedro M},
  journal={Journal of the Mechanics and Physics of Solids},
  volume={61},
  number={8},
  pages={1712--1736},
  year={2013},
  publisher={Elsevier},
  doi={10.1016/j.jmps.2013.04.002}
}

@article{lazarus2013contorting,
  title={Contorting a heavy and naturally curved elastic rod},
  author={Lazarus, Arnaud and Miller, Jay T and Metlitz, Matthew M and Reis, Pedro M},
  journal={Soft Matter},
  volume={9},
  number={34},
  pages={8274--8281},
  year={2013},
  publisher={Royal Society of Chemistry},
  doi={10.1039/c3sm50873k}
}

@article{coyne1990analysis,
  title={Analysis of the formation and elimination of loops in twisted cable},
  author={Coyne, James},
  journal={IEEE Journal of Oceanic Engineering},
  volume={15},
  number={2},
  pages={72--83},
  year={1990},
  publisher={IEEE},
  doi={10.1109/48.50692}
}

@article{thompson1996helix,
  title={From helix to localized writhing in the torsional post-buckling of elastic rods},
  author={Thompson, John Michael Tutill and Champneys, AR},
  journal={Proceedings of the Royal Society of London. Series A: Mathematical, Physical and Engineering Sciences},
  volume={452},
  number={1944},
  pages={117--138},
  year={1996},
  publisher={The Royal Society London},
  doi={10.1098/rspa.1996.0007}
}

@article{goriely1998nonlinear,
  title={Nonlinear dynamics of filaments. IV Spontaneous looping of twisted elastic rods},
  author={Goriely, Alain and Tabor, Michael},
  journal={Proceedings of the Royal Society of London. Series A: Mathematical, Physical and Engineering Sciences},
  volume={454},
  number={1980},
  pages={3183--3202},
  year={1998},
  publisher={The Royal Society},
  doi={10.1098/rspa.1998.0297}
}

@article{petruska2012optimal,
  title={Optimal permanent-magnet geometries for dipole field approximation},
  author={Petruska, Andrew J and Abbott, Jake J},
  journal={IEEE transactions on magnetics},
  volume={49},
  number={2},
  pages={811--819},
  year={2012},
  publisher={IEEE},
  doi={10.1109/tmag.2012.2205014}
}

@article{thompson2022lammps,
  title={{LAMMPS}-a flexible simulation tool for particle-based materials modeling at the atomic, meso, and continuum scales},
  author={Thompson, Aidan P and Aktulga, H Metin and Berger, Richard and Bolintineanu, Dan S and Brown, W Michael and Crozier, Paul S and In't Veld, Pieter J and Kohlmeyer, Axel and Moore, Stan G and Nguyen, Trung Dac and others},
  journal={Computer physics communications},
  volume={271},
  pages={108171},
  year={2022},
  publisher={Elsevier},
  doi={10.1016/j.cpc.2021.108171}
}

@article{denniston2022lammps,
  title={{LAMMPS} lb/fluid fix version 2: Improved hydrodynamic forces implemented into {LAMMPS} through a lattice-{Boltzmann} fluid},
  author={Denniston, Colin and Afrasiabian, Navid and Cole-Andr{\'e}, MG and Mackay, FE},
  journal={Computer Physics Communications},
  volume={275},
  pages={108318},
  year={2022},
  publisher={Elsevier},
  doi={10.1016/j.cpc.2022.108318}
}

@article{tan2018parallel,
  title={A parallel fluid--solid coupling model using {LAMMPS} and {Palabos} based on the immersed boundary method},
  author={Tan, Jifu and Sinno, Talid R and Diamond, Scott L},
  journal={Journal of Computational Science},
  volume={25},
  pages={89--100},
  year={2018},
  publisher={Elsevier},
  doi={https://doi.org/10.1016/j.jocs.2018.02.006}
}

@article{ye2020openfsi,
  title={{OpenFSI}: A highly efficient and portable fluid--structure simulation package based on immersed-boundary method},
  author={Ye, Huilin and Shen, Zhiqiang and Xian, Weikang and Zhang, Teng and Tang, Shan and Li, Ying},
  journal={Computer Physics Communications},
  volume={256},
  pages={107463},
  year={2020},
  publisher={Elsevier},
  doi={10.1016/j.cpc.2020.107463}
}

@article{bhatnagar1954model,
  title={A model for collision processes in gases. {I}. Small amplitude processes in charged and neutral one-component systems},
  author={Bhatnagar, Prabhu Lal and Gross, Eugene P and Krook, Max},
  journal={Physical review},
  volume={94},
  number={3},
  pages={511},
  year={1954},
  publisher={APS},
  doi={10.1103/physrev.94.511}
}

@book{kruger2017lattice,
  title={The lattice {Boltzmann} method},
  author={Kr{\"u}ger, Timm and Kusumaatmaja, Halim and Kuzmin, Alexandr and Shardt, Orest and Silva, Goncalo and Viggen, Erlend Magnus},
  volume={10},
  number={978-3},
  year={2017},
  isbn={978-3-319-44649-3},
  publisher={Springer},
  address={Switzerland},
  doi={10.1007/978-3-319-44649-3}
}

@article{guo2002discrete,
  title={Discrete lattice effects on the forcing term in the lattice {Boltzmann} method},
  author={Guo, Zhaoli and Zheng, Chuguang and Shi, Baochang},
  journal={Physical Review E},
  volume={65},
  number={4},
  pages={046308},
  year={2002},
  publisher={APS},
  doi={10.1103/physreve.65.046308}
}

@article{peskin2002immersed,
  title={The immersed boundary method},
  author={Peskin, Charles S},
  journal={Acta numerica},
  volume={11},
  pages={479--517},
  year={2002},
  publisher={Cambridge University Press},
  doi={10.1017/s0962492902000077}
}

@article{tretyakov2017improved,
  title={An improved dissipative coupling scheme for a system of Molecular Dynamics particles interacting with a Lattice {Boltzmann} fluid},
  author={Tretyakov, Nikita and D{\"u}nweg, Burkhard},
  journal={Computer Physics Communications},
  volume={216},
  pages={102--108},
  year={2017},
  publisher={Elsevier},
  doi={10.1016/j.cpc.2017.03.009}
}

@article{vlogman2024efficient,
  title={Efficient coupled lattice {Boltzmann} and Discrete Element Method simulations of small particles in complex geometries},
  author={Vlogman, Tristan G and Jain, Kartik},
  journal={Computers \& Mathematics with Applications},
  volume={175},
  pages={313--329},
  year={2024},
  publisher={Elsevier},
  doi={10.1016/j.camwa.2024.10.004}
}

@article{schunter2018elastogranular,
  title={Elastogranular mechanics: buckling, jamming, and structure formation},
  author={Schunter Jr, David J and Brandenbourger, Martin and Perriseau, Sophia and Holmes, Douglas P},
  journal={Physical Review Letters},
  volume={120},
  number={7},
  pages={078002},
  year={2018},
  publisher={APS},
  doi={10.1103/physrevlett.120.078002}
}

@article{manning2023essay,
  title={Essay: Collections of deformable particles present exciting challenges for soft matter and biological physics},
  author={Manning, M Lisa},
  journal={Physical Review Letters},
  volume={130},
  number={13},
  pages={130002},
  year={2023},
  publisher={APS},
  doi={10.1103/physrevlett.130.130002}
}

@article{pinelli2017pelskin,
  title={The {PELskin} project: part {IV}—control of bluff body wakes using hairy filaments},
  author={Pinelli, Alfredo and Omidyeganeh, Mohammad and Br{\"u}cker, Christoph and Revell, Alistair and Sarkar, Abhishek and Alinovi, Edoardo},
  journal={Meccanica},
  volume={52},
  number={7},
  pages={1503--1514},
  year={2017},
  publisher={Springer},
  doi={10.1007/s11012-016-0513-0}
}

@article{agrawal2024efficient,
  title={An efficient isogeometric/finite-difference immersed boundary method for the fluid--structure interactions of slender flexible structures},
  author={Agrawal, Vishal and Kulachenko, Artem and Scapin, Nicol{\`o} and Tammisola, Outi and Brandt, Luca},
  journal={Computer Methods in Applied Mechanics and Engineering},
  volume={418},
  pages={116495},
  year={2024},
  publisher={Elsevier},
  doi={10.1016/j.cma.2023.116495}
}

@article{jiang2023numerical,
  title={Numerical Study of Metachronal Wave-Modulated Locomotion in Magnetic Cilia Carpets},
  author={Jiang, Hao and Gu, Hongri and Nelson, Bradley J and Zhang, Teng},
  journal={Advanced Intelligent Systems},
  volume={5},
  number={10},
  pages={2300212},
  year={2023},
  publisher={Wiley Online Library},
  doi={10.1002/aisy.202300212}
}

@article{huang2023modeling,
  title={Modeling of magnetic cilia carpet robots using discrete differential geometry formulation},
  author={Huang, Weicheng and Liu, Mingchao and Hsia, K Jimmy},
  journal={Extreme Mechanics Letters},
  volume={59},
  pages={101967},
  year={2023},
  publisher={Elsevier},
  doi={10.1016/j.eml.2023.101967}
}

@misc{swing-twist,
    author = {Baker, M. J.},
    title = {Maths - Decomposition of rotations.},
    howpublished = {\url{https://www.euclideanspace.com/maths/geometry/rotations/for/decomposition/index.htm}},
    note = {Accessed: 2025-12-02}
}

@article{hackney2026shape,
  author ="Hackney, Nicholas W. and Clemmer, Joel T. and Grest, Gary S.",
  title  ="Shape elasticity in colloidal bent-core liquid crystals",
  journal  ="Soft Matter",
  year  ="2026",
  volume  ="22",
  issue  ="16",
  pages  ="3126-3134",
  publisher  ="The Royal Society of Chemistry",
  doi  ="10.1039/D6SM00016A"
}

@article{pezeshki2025tunable,
  title={Tunable entanglement and strength with engineered staple-like particles: Experiments and discrete element models},
  author={Pezeshki, Saeed and Sohn, Youhan and Fouquet, Vivien and Barthelat, Francois},
  journal={Journal of the Mechanics and Physics of Solids},
  volume={200},
  pages={106127},
  year={2025},
  publisher={Elsevier},
  doi={10.1016/j.jmps.2025.106127}
}

@article{pol2025granular,
  title={Granular drag and lift force on a flexible fiber},
  author={Pol, Antonio and Storti, Sara and Gabrieli, Fabio},
  journal={Physical Review E},
  volume={112},
  number={4},
  pages={045425},
  year={2025},
  publisher={APS},
  doi={10.1103/p2db-yt7h}
}

\section*{Figures}

\textbf{Fig.~\ref{fig:magnetic-bpm-intro} Illustration of the bonded-particle model for magneto-elastic rods.} 
(a)~Each particle is treated as a magnetic dipole. An external magnetic field or dipole--dipole interaction exerts a force and torque on the particle. Neighboring particles are connected by an elastic bond. 
(b)~The relative displacement and rotation between two bonded particles are decomposed into four linear spring interactions. The reference configurations are displayed below the corresponding deformed configurations.

\noindent
\textbf{Fig.~\ref{fig:BPM-frames} Definition of the BPM frames.} 
(a)~The reference (light) and deformed (dark) bond. 
(b)~The body frames of the two particles are defined by the quaternions $q_1$ and $q_2$ in the global frame. The central frame, $C$, is defined by $q_C$, which is the average of $q_1$ and $q_2$. 
(c)~The relative displacement is measured from the moving $C$ frame. 
(d)~The $C'$ frame is obtained by aligning the $C$ frame to the bond vector. 
(e)~The orientation of particles $1$ and $2$ in the $C'$ frame is defined by the quaternions $u^\ast$ and $u$, respectively. Both quaternions are then decomposed in a \textit{bend before twist} order such that the total rotation of particle $2$ with respect to particle $1$ is a sequence of four rotations. 
(f)~Illustration of the swing--twist decomposition.

\noindent
\textbf{Fig.~\ref{fig:heavy-rods-setup} The setup for the twisting experiment of \citet{lazarus2013continuation}.} 
(a)~An initially straight rod is compressed by a fixed amount causing it to buckle under its own weight. The rod is then twisted multiple times and the resulting shape is studied (Fig.~\ref{fig:heavy-straight-rod}). 
(b)~An ideal naturally curved rod coiled into a circle with curvature $\kappa$ can be straightened by applying a pure end moment $M = \kappa E I $. The straightened rod then undergoes the same twisting procedure as in (a) (Fig.~\ref{fig:heavy-curved-rod}).

\noindent
\textbf{Fig.~\ref{fig:heavy-straight-rod} Extreme twisting of a pre-buckled initially straight rod.} 
(a)~Top view snapshots at different twisting angles $\Phi$ comparing the experimental images of \citet{lazarus2013continuation} and ours. 
(b)~The maximum deflection as a function of the twist angle. The red points are their experimental measurements and the green curve is their semi-analytical solution, while the blue curve is our simulation result. The simulated plectoneme (inset) formation happens near their predicted critical angle $\theta_c = 11.33\pi$.

\noindent
\textbf{Fig.~\ref{fig:heavy-curved-rod} Extreme twisting of a pre-buckled, naturally curved rod.} 
The rod is coiled in its stress-free state and straightened before undergoing the same procedure as in Fig.~\ref{fig:heavy-straight-rod}. 
(a)~Top view snapshots at different twisting angles comparing the experimental images of \citet{lazarus2013continuation} with ours. 
(b)~Maximum deflection as a function of twist angle. The red points are their experimental measurements and the green curve is their semi-analytical solution, while the blue curve is our simulation result. The plectoneme forms at a critical angle $\theta_c = 16.15 \pi$.

\noindent
\textbf{Fig.~\ref{fig:magnetic-beams} Magnetized beams under uniform and constant-gradient fields.} 
The schematics in the left column were taken from \citet{yan_comprehensive_2022}. 
(a)~The beam is magnetized in the $+y$ direction and $(\nabla \bm{B})_{yy} > 0$. The $x$ and $y$ components of the tip displacement are compared with FEM results. 
(b)~The tip deflection in the $y$ direction compared with FEM results for three different configurations.

\noindent
\textbf{Fig.~\ref{fig:magnetic-helix} Magnetic helical rods with dipole--dipole interactions.} 
(a)~Simulation snapshots for the helix with $\gamma = 4 \times 10^{-4}$. The top row shows notable configurations for increasing $\lambda_m$, and the bottom row for decreasing $\lambda_m$. 
(b)~The normalized distance as a function of $\lambda_m$ (external field strength) for different values of $\gamma$ (dipole--dipole interaction strength).

\noindent
\textbf{Fig.~\ref{fig:filaments-oscillating-flow} Filaments in an oscillatory flow.} 
(a)~Simulation setup. 
(b)~Displacement of the rightmost filament's tip (highlighted in red in (a)) over successive cycles, compared with the results of \citet{pinelli2017pelskin}. 
(c)~The out-of-plane component of the vorticity in the plane containing the filaments during the maximum and minimum displacement. The scale bar (yellow) is $L = 10~\unit{mm}$.

\noindent
\textbf{Fig.~\ref{fig:magnetic-cilia-array} Fluid pumping by magnetic cilia arrays.} 
(a)~Side view of the cilia array and representative magnetization patterns, controlled by the wavelength $\lambda$. Each row of cilia (same $y$, into the page) has the same magnetization, whose direction is indicated by the corresponding arrow in the table. The plate (red particles) lies in the $xy$-plane and is not included in the simulation, serving only for visualization. 
(b)~The cilia array is immersed in an open channel. A rotating uniform magnetic field is applied to actuate the cilia, and tracer particles are added to visualize the flow pattern. 
(c)~Comparison of tracer trajectories between experimental images of \citet{Gu2020} and our simulations. 
(d)~The average net displacement per cycle at various heights above the cilia tip. The colors correspond to those in (c). The inset shows the average displacement for the region $[0, 4~\unit{mm}]$ above the cilia tip.

\noindent
\textbf{Fig.~\ref{fig:LBM-BPM} Schematic of the LBM--BPM coupling.} 
A rod is immersed in a fluid grid. Each LBM step updates the macroscopic fluid properties by collision and streaming of the distribution functions. For each DEM (sub)step, the fluid velocity is interpolated to the current position of the particle. A penalty force is computed from the difference between the interpolated fluid velocity and the actual particle velocity, where equal and opposite forces are applied to the particle and spread into the fluid grid.

\end{document}


\maketitle
\tableofcontents

\section{LAMMPS implementation}\label{sec:LAMMPS}

The LAMMPS input scripts and data files for all examples are available at \url{https://gsalkuin.github.io/lammps-bpm-rods/}.
The input script is written in a domain-specific language and contains all the instructions to run the simulation from start to finish.
Each line is a \textit{command}, whose details can be found in the official documentation (\url{https://docs.lammps.org/}). 
Here, we use \texttt{monospace font} to denote LAMMPS commands/styles. 
Some functionalities require additional \textit{packages} to be enabled before building LAMMPS. 
In this work, we use the following packages: BPM, DIPOLE, GRANULAR, LEPTON, MOLECULE.

\subsection{BPM package}\label{sec:BPM-pkg}

The BPM package was developed by Clemmer et al.~\citep{clemmer2024soft}, providing the framework on which the present model is built.
The \texttt{atom\_style bpm/sphere} defines spherical particles with orientation represented by a unit quaternion. 
BPM particles are integrated in time using velocity Verlet for translations and Richardson extrapolation for quaternion orientations, as implemented in \texttt{fix nve/bpm/sphere}. 
The \texttt{update dipole} keyword--value pair can also be added to rotate the dipole moments with the quaternions.
The elastic rod model developed in this paper has been merged into \texttt{bond\_style bpm/rotational}, which is available in LAMMPS as of the 4 July 2026 feature release (\url{https://docs.lammps.org/bond_bpm_rotational.html}).
A bond is defined using a bond ID, a bond type, and two atom IDs. Bonds can be defined in the data file or in the input script using \texttt{create\_bonds}.

For the examples with contact, \texttt{pair\_style granular} was used.
To create a smooth contact surface and prevent other particles from passing through the gap between bonded particles, the bond length is set shorter than the particle diameter, resulting in overlap between bonded particles. 
In that case, the pair interaction can be disabled for particles connected up to three bonds away using the command \texttt{special\_bonds lj 0 0 0}.

\subsubsection*{Equations of motion}\label{sec:EOM}

Each particle $\alpha$ stores a mass $m_\alpha$, moment of inertia $I_\alpha = \frac{2}{5} m_\alpha R_\alpha^2$ (solid sphere), position $\bm{x}_\alpha$, velocity $\bm{v}_\alpha$, unit quaternion $q_\alpha$, and angular velocity $\bm{\omega}_\alpha$. 
There is also an option to use a disc for 2D by adding the \texttt{disc} keyword. 
The equations of motion are
\begin{align*}
  m_\alpha \dot{\bm{v}}_\alpha &= \bm{f}_\alpha, \\
  I_\alpha \dot{\bm{\omega}}_\alpha &= \bm{\tau}_\alpha,
\end{align*}
where $\bm{f}_\alpha$ and $\bm{\tau}_\alpha$ are the total force and torque on particle $\alpha$, comprising contributions from elasticity (Eq.~\eqref{M-eq:bpm-vector}), bond and global damping, granular contact, magnetic interactions, and other external forces.

In \texttt{fix nve/bpm/sphere}, $\bm{x}$, $\bm{v}$, and $\bm{\omega}$ are integrated using velocity Verlet, 
and $q$ via Richardson extrapolation.
Given $\bm{f}$ and $\bm{\tau}$ at time $t$, the half-step update is:
\begin{align*}
  \bm{v}(t+\Delta t/2) &= \bm{v}(t)
    + \frac{\Delta t}{2} \frac{\bm{f}(t)}{m}, \\
  \bm{x}(t+\Delta t) &= \bm{x}(t)
    + \Delta t \, \bm{v}(t+\Delta t/2), \\
  \bm{\omega}(t+\Delta t/2) &= \bm{\omega}(t)
    + \frac{\Delta t}{2} \frac{\bm{\tau}(t)}{I}.
\end{align*}
The quaternion is updated using Richardson extrapolation.
Let $\omega = (0, \bm{\omega})$ and $\dot{q} = \frac{1}{2}\omega\,q$.
Two approximations to $q(t + \Delta t)$ are computed, both using $\bm{\omega}(t+\Delta t/2)$.
The first uses a single step of size $\Delta t$:
\begin{align*}
  q_{(1)} &= q + \Delta t\,\dot{q}.
\end{align*}
The second uses two successive steps of size $\Delta t / 2$:
\begin{align*}
  q' &= q + \frac{\Delta t}{2}\,\dot{q}, \\
  q_{(2)} &= q' + \frac{\Delta t}{2}\,\dot{q}'.
\end{align*}
Then the following has an error quadratic in $\Delta t$:
\begin{align*}
  q(t+\Delta t) &= 2\,q_{(2)} - q_{(1)}.
\end{align*}
All quaternions are renormalized to unit length after each substep.
After recomputing $\bm{f}$ and $\bm{\tau}$ at $t+\Delta t$, the velocities are updated:
\begin{align*}
  \bm{v}(t+\Delta t) &= \bm{v}(t+\Delta t/2)
    + \frac{\Delta t}{2} \frac{\bm{f}(t+\Delta t)}{m}, \\
  \bm{\omega}(t+\Delta t) &= \bm{\omega}(t+\Delta t/2)
    + \frac{\Delta t}{2} \frac{\bm{\tau}(t+\Delta t)}{I}.
\end{align*}

\subsection{Magnetic dipoles}

The \texttt{fix efield} command is used to apply a uniform magnetic field, and the \texttt{fix efield/lepton} command is used for a non-uniform magnetic field. 
Magnetic dipole--dipole interaction can be added using \texttt{pair\_style lj/cut/dipole/cut} or other variants. 
To rotate the dipole with the particle, as required by the magnetic model used in this work \citep{danas_stretch-independent_2024}, the \texttt{update dipole} keyword--value pair must be added to \texttt{fix nve/bpm/sphere}.
In this work, we only consider permanent dipoles, but other magnetization models can also be used. 
For example, a simple superparamagnetic model can be implemented using the \texttt{fix set} command to update the dipoles directly together with \texttt{pair\_style lj/cut/dipole/cut}. 
More complicated magnetization models may require users to modify the LAMMPS source code.

\subsection{Unit systems}

With magnetic dipoles, it is best to use \texttt{units lj} or \texttt{units cgs}. 
Since the \texttt{efield} commands were made for electric dipoles, other unit systems may have internal conversion factors that users should be aware of. 
For example, in \texttt{units si} the dipole--dipole energy will have a prefactor $(4\pi\varepsilon_0)^{-1} = 8.9876\times10^9~\unit{N}\,\unit{m}^2/\unit{C}^2$ (Coulomb constant), so the consistent magnetic units are $\mu^\ast = c \, [\unit{C}\,\unit{m}] = 3\times10^8 \,[\unit{A}\,\unit{m}^2]$ and $B^\ast = c^{-1} \, [\unit{V}/\unit{m}] = 3.33\times10^{-9} \,[\unit{T}]$.

LAMMPS' \texttt{cgs} units assume ESU, which is equivalent to EMU for purely magnetostatic problems.
The \textit{milli} system is not built into LAMMPS; here we define it such that $\mu_0 = 4\pi$.
Its temperature scale is set by $\{k_B\}$, the numerical value of the Boltzmann constant in the chosen LAMMPS unit system.
In \texttt{lj}, $\{k_B\} = 1$, while in \texttt{cgs}, $\{k_B\} = 1.38 \times 10^{-16}$.

\begin{table*}[htbp]
\centering
\caption{Unit systems used in the LAMMPS simulations.}
\label{tab:units}
\begin{minipage}[t]{0.48\textwidth}
\centering
\subcaption{CGS (EMU)}
\label{tab:units-cgs}
\begin{tabular}{lll}
\hline
\textbf{Quantity} & \textbf{Unit} & \textbf{SI} \\
\hline
Length              & cm              & $10^{-2}$\,m           \\
Time                & s               & $1$\,s                 \\
Mass                & g               & $10^{-3}$\,kg          \\
Force               & dyne            & $10^{-5}$\,N           \\
Energy              & erg             & $10^{-7}$\,J           \\
Density             & g/cm$^3$        & $10^{3}$\,kg/m$^3$     \\
Stress              & Ba              & $10^{-1}$\,Pa          \\
Mag.\ dipole        & emu             & $10^{-3}$\,A\,m$^2$    \\
Mag.\ field         & G               & $10^{-4}$\,T           \\
Kin.\ viscosity     & cm$^2$/s        & $10^{-4}$\,m$^2$/s     \\
Temperature         & K               & $1$\,K                 \\
\hline
\end{tabular}
\end{minipage}
\hfill
\begin{minipage}[t]{0.48\textwidth}
\centering
\subcaption{milli (mm, ms, mg, $\mu_0=4\pi$)}
\label{tab:units-milli}
\begin{tabular}{lll}
\hline
\textbf{Quantity} & \textbf{Unit} & \textbf{SI} \\
\hline
Length              & mm              & $10^{-3}$\,m           \\
Time                & ms              & $10^{-3}$\,s           \\
Mass                & mg              & $10^{-6}$\,kg          \\
Force               & mN              & $10^{-3}$\,N           \\
Energy              & $\mu$J          & $10^{-6}$\,J           \\
Density             & mg/mm$^3$       & $10^{3}$\,kg/m$^3$     \\
Stress              & kPa             & $10^{3}$\,Pa           \\
Mag.\ dipole        & --              & $10^{-4}$\,A\,m$^2$    \\
Mag.\ field         & --              & $10^{-2}$\,T           \\
Kin.\ viscosity     & mm$^2$/ms       & $10^{-3}$\,m$^2$/s     \\
Temperature         & --              & $10^{-6}/\{k_B\}$\,K    \\
\hline
\end{tabular}
\end{minipage}
\end{table*}

\section{BPM formulation details}\label{sec:moreBPM}

\subsection{Symmetric rotational decomposition}\label{sec:sym-rot}

In the BPM of Wang et al.~\citep{wang_implementation_2006, wang2008macroscopic, wang2009new, wang_esys_particle_2009} (and most beam-bond models \citep{chen_comparative_2022}), one particle is chosen as the reference, from which the motion of the other particle is decomposed. 
However, we found that for non-planar deformations, the decomposition algorithm of \citet{wang2009new} for finite rotation is not permutation invariant; i.e., for a given bond, different forces and torques will be computed depending on which particle was chosen as the reference. 
This is because the non-commutativity of rotations makes it impossible to prescribe the same rotation order for both particles, leading to different bending planes and therefore different bending and rotational shear terms. 
This reference frame dependence is not unique to the formulation of \citet{wang2009new}; co-rotational models generally require specification of a distinguished frame, and many choices are possible \citep{crisfield1990consistent,obermayr2013bonded}. 
We modify the model by introducing a \textit{central frame} obtained by averaging the two particles' orientations using quaternion spherical linear interpolation, similar to the model of \citet{obermayr2013bonded}. 
Measuring the shear in this frame also combines the otherwise separate translational and rotational shear components into one translational shear term.

\begin{figure}[!htpb]
    \centering
    \includegraphics[width=0.75\textwidth]{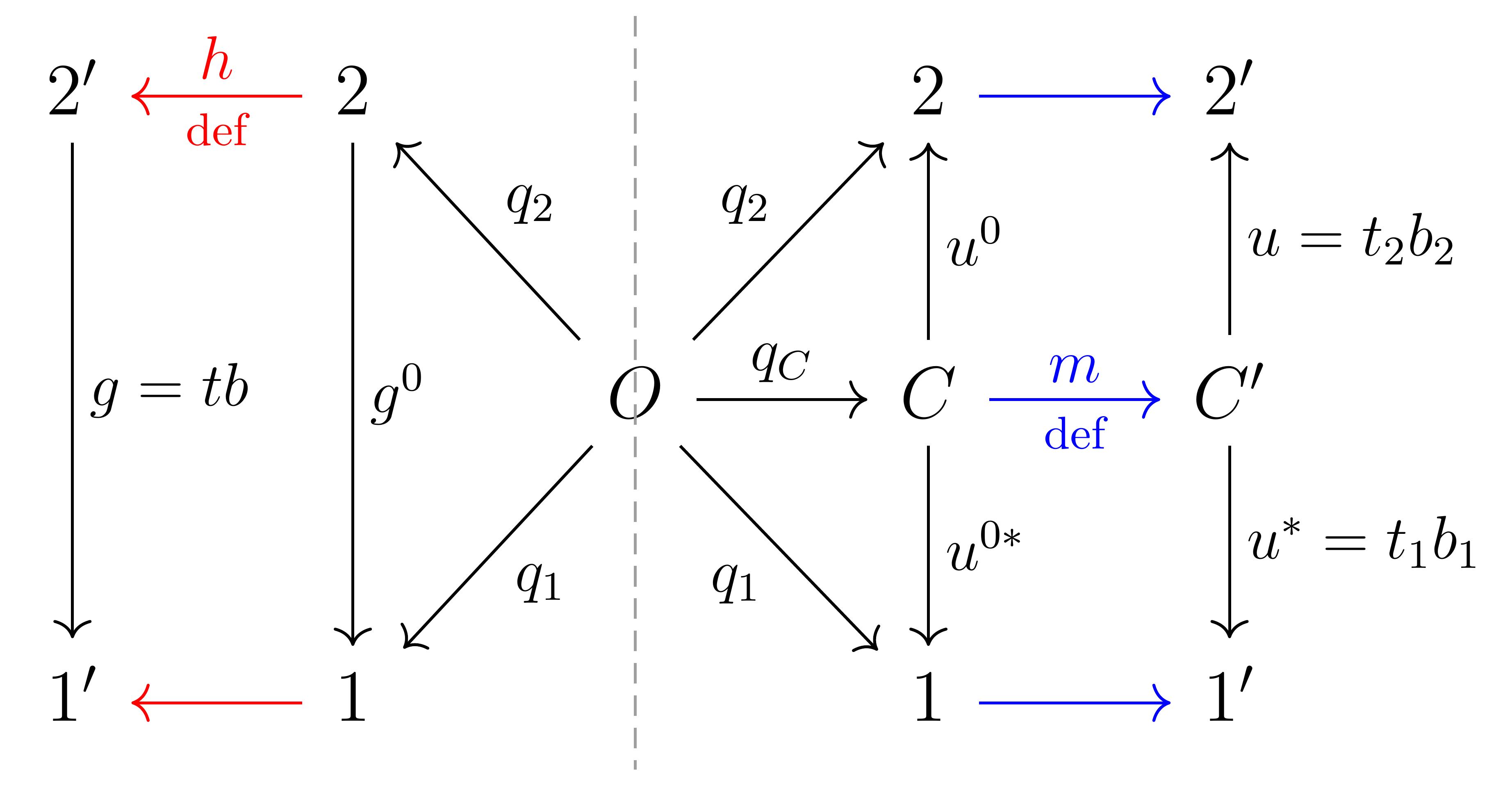}
    \caption{The different reference frames in (left of $O$) the original \citep{wang_implementation_2006, wang2008macroscopic, wang2009new, wang_esys_particle_2009} and (right of $O$) the modified BPM formulation. All frames (except $O$) are in the current configuration. A quaternion is represented by an arrow, where the base is the reference frame in which the rotation is defined. A quaternion from $O$ gives the orientation of the particle's body-fixed frame. Some properties: (1) A rotation $p: A \to B$, followed by another $q: B \to C$ is equivalent to the quaternion $p q: A \to C$ (extrinsic). (2)~If a vector $\bm{v}$ is defined on frame $A$ and $q: A \to B$, then the vector becomes $q^\ast \bm{v} q$ in frame $B$. The central frame $C$ is obtained by averaging the two particle frames. The frame $C'$ is obtained by rotating $C$ with $m$ such that its $z$-axis is aligned to $\bm{r}_f[C]$. A change in reference frame is obtained by rotating $1$, $2$, and $C$ by $m$. The quaternion $u$ rotates $C'$ to $2'$ and its conjugate rotates $C'$ to $1'$. Both $u$ and $u^\ast$ are decomposed using a \textit{bend before twist} decomposition. (Since the quaternions $b_i$ and $t_i$ are defined in the $C'$ frame, their composition order is right to left.) We then introduce the effective ``bending'' $b = b^{}_2 b^\ast_1$ and ``twisting'' $t = t^{}_2 t^\ast_1$ rotations defined in the $C'$ frame.}
    \label{fig:BPM-symmetric-reference-frames}
\end{figure}

The asymmetry in the original BPM formulation can be seen in Fig.~\ref{fig:BPM-symmetric-reference-frames} (left of $O$). 
Exchanging $1$ and $2$ leads to different $h$, $t$, and $b$ quaternions. The introduction of the $C$ frame (Fig.~\ref{fig:BPM-symmetric-reference-frames}, right of $O$) removes the preference of one particle over the other. 
The $C'$ frame is defined by rotating $C$ such that its $z$-axis aligns with the bond vector. 
This makes it such that in the $C'$ frame the twist axis is along $z$. The frame rotation is described by the quaternion $m$, and its axis is parallel to $\hat{\bm{z}}[C] \times \bm{r}_f[C]$. 
Let $\bm{r}_f[C] = \langle x, y, z \rangle$ and $r = ||\bm{r}_f||$. Then, from Eq.~20 in \citet{wang2009new},
\begin{align}\label{eq:quat-h}
    m_0 &= \frac{\sqrt{2}}{2}\sqrt{1+\frac{z}{r}}, \nonumber\\
    m_1 &= -\frac{\sqrt{2}}{2}\sqrt{1-\frac{z}{r}} \frac{y}{\sqrt{x^2 +y^2}}, \nonumber\\
    m_2 &= \frac{\sqrt{2}}{2}\sqrt{1-\frac{z}{r}} \frac{x}{\sqrt{x^2 +y^2}}, \nonumber\\
    m_3 &= 0 .
\end{align}
The procedure of constructing the $C'$ frame is similar to \citet{crisfield1990consistent,obermayr2013bonded}; however, in \citet{crisfield1990consistent} this was only approximated using a mid-point rule. 

Since $q_C$ is the average of $q_1$ and $q_2$, we define the half-rotation $u^0_{(C \to 2) \,=\, (1 \to C)} = q_C^\ast q_2 = q_1^\ast q_C$. 
In the $C'$ frame, this becomes $u = m^\ast u^0 m$. 
The total rotation from $1'$ to $2'$ is $u \, u = u^2 = m^\ast q_1^\ast q_2 m$. 
However, rather than decomposing this rotation into bend and twist, we decompose each half-rotation separately: $u = t_2 b_2$ and $u^\ast = t_1 b_1$. 
This yields a symmetric $t_2 b_2 b_1^\ast t_1^\ast$ sequence in $C'$. 
From this, we define the effective ``bending'' $b = b^{}_2 b^\ast_1$ and ``twisting'' $t = t^{}_2 t^\ast_1$. 
Combining the two bends into one guarantees angular momentum conservation. 
Exchanging $1$ and $2$ leads to the new $m$ being related to the old $m$ by a $180^\circ$ rotation about any axis perpendicular to $z$, which only swaps $u$ and $u^\ast$, guaranteeing symmetry.

Note that $t_2 b_2 b_1^\ast t_1^\ast = b_1^\ast t_1^\ast t_2 b_2$ since the total rotation is $u u$. 
This implies that the effective bending could be defined as either $b = b_2 b_1^\ast$ or $b = b_1^\ast b_2$; however, these are not equivalent since rotations do not commute. 
Moreover, any rotation can be parametrized by three parameters. 
Both $t_1$ and $t_2$ rotate about $\hat{z}$, yielding a single twist parameter $\psi_2 - \psi_1$. 
However, the two bending rotations $b_1$ and $b_2$ generally occur in different planes. 
This gives a total of five parameters, which overdetermines the rotation. 
The chosen decomposition ensures consistency under particle exchange; whether there are better alternative parametrizations that preserve this symmetry remains an open question.

\subsection{Deriving the spring constants}\label{sec:bpm-constants}

For the axial stretching and twisting, we assume that the bond connecting two particles is a cylinder of length $L$ with a uniform circular cross section of area $A$. 
The case with non-uniform cross section is also considered in \citet{wang_esys_particle_2009} by scaling the torques due to the movement of the ``contact point''; however, we do not consider that here. 
The radial spring stiffness $K_r$ can be derived from the definition of Young's modulus: $E = (\text{tensile stress})/ (\text{tensile strain}) \approx (F/A)\big/(\Delta L/L)$. 
Since $F \approx K_r \Delta L$, we obtain $K_r = EA/L$. 
The case for the twisting stiffness is similar. 
Here, we use the well-known result from the torsion of beams to obtain $K_t = GJ/L$ (see \citet{slaughter2012linearized}, Ch.~1.3), where $J$ is the polar second moment of area.

\begin{figure}[!htpb]
    \centering
    \includegraphics[width=0.6\textwidth]{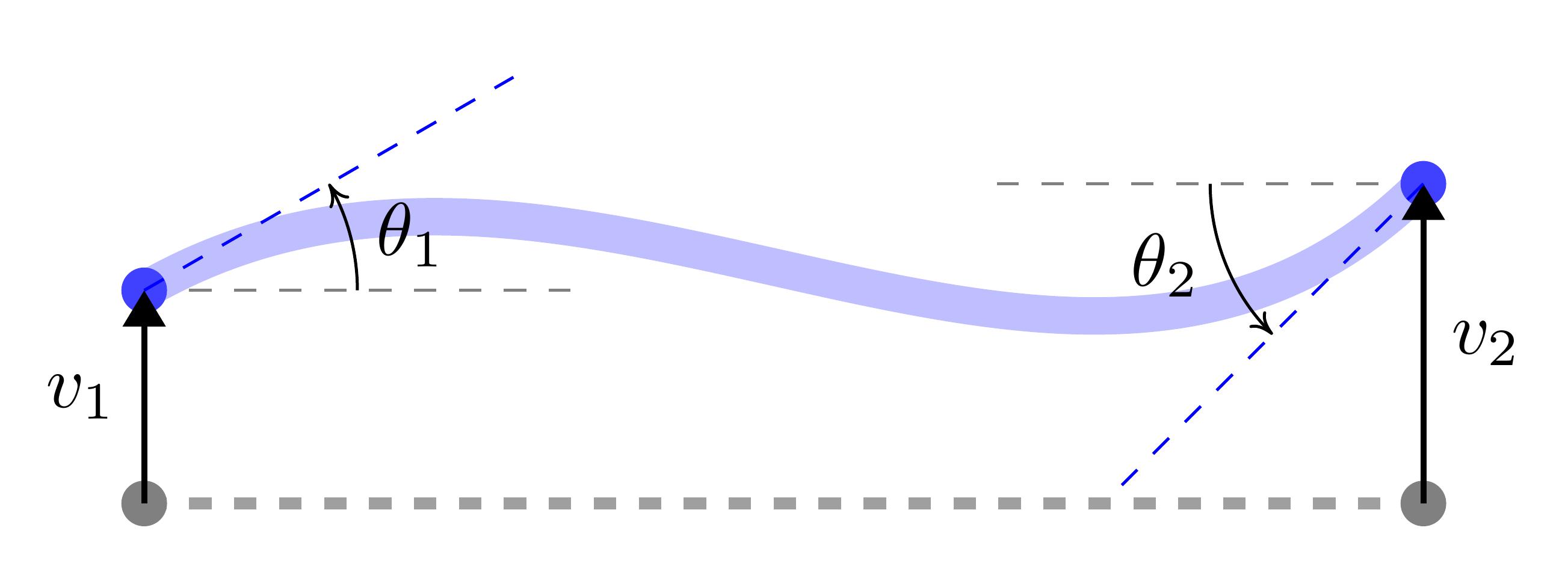}
    \caption{Local frame of the beam element. Positive displacement and force are along the vertical direction. Positive angle and applied torque are counter-clockwise.}
    \label{fig:beam-element}
\end{figure}

Now consider the case of planar bending. 
For the moment, we assume that there is neither stretching nor twisting, which we treat separately since we linearize the forces and torques. 
We base our model on the Euler--Bernoulli (EB) beam theory, which assumes unshearable beams. 
In this case, the ``shear'' is actually due to transverse displacements. 
For the two-node beam finite element shown in Fig.~\ref{fig:beam-element}, using cubic Hermite shape functions, the following matrix equation can be obtained (see \citet{reddy1993introduction, hutton2003FEA} for more details):
\begin{equation}\label{eq:beam-stiffness-matrix}
    \begin{bmatrix}
    f_1 \\
    \tau_1 \\
    f_2 \\
    \tau_2 
    \end{bmatrix} = -\frac{EI}{L^3}
    \begin{bmatrix}
    12 & 6L & -12 & 6L \\
    6L & 4L^2 & -6L & 2L^2 \\
    -12 & -6L & 12 & -6L \\
    6L & 2L^2 & -6L & 4L^2
    \end{bmatrix}
    \begin{bmatrix}
    v_1 \\
    \theta_1 \\
    v_2 \\
    \theta_2
    \end{bmatrix}
\end{equation}
The negative sign is due to the fact that in the BPM, $f$ and $\tau$ represent the restoring force and torque, rather than the applied ones. 
In matrix form, $\bm{f} = - \mathbf{K} \bm{u}_g$. 
The null space of $\mathbf{K}$ is spanned by $\bm{a} = (1,0,1,0)$ and $\bm{b} = (-L/2,1,L/2,1)$; any linear combination of these vectors added to $\bm{u}_g$ will not change $\bm{f}$ and can be interpreted as a change of reference frame.

For example, the generalized displacement vector in the central frame is $\bm{u}_c := \bm{u}_g - \bar{v} \bm{a} - \bar{\theta} \bm{b}$. 
Then, $\bm{f} = - \mathbf{K} \bm{u}_c$ becomes
\begin{equation}\label{eq:bpm-general-planar}
    \begin{bmatrix}
    f_1 \\
    \tau_1 \\
    f_2 \\
    \tau_2 
    \end{bmatrix} \approx \frac{EI}{L^3}
    \begin{bmatrix}
    12 (v_2 - v_1 - L \bar{\theta}) \\
    6 L (v_2 - v_1 - L \bar{\theta}) + L^2 (\theta_2 - \theta_1) \\
    - 12 (v_2 - v_1 - L \bar{\theta}) \\
    6 L (v_2 - v_1 - L \bar{\theta}) - L^2 (\theta_2 - \theta_1)
    \end{bmatrix} =
    \begin{bmatrix}
    K_s \delta s \\
    K_s \delta s L/2 + K_b \theta_b \\
    - K_s \delta s \\
    K_s \delta s L/2 - K_b \theta_b
    \end{bmatrix} \approx
    \begin{bmatrix}
    f_s \\
    \tau_s + \tau_b \\
    - f_s \\
    \tau_s - \tau_b
    \end{bmatrix},
\end{equation}
where $\delta s := v_2 - v_1 - L \bar{\theta}$, $\bar{\theta} := (\theta_1+\theta_2)/2$, and $\theta_b := \theta_2 - \theta_1$. 
In the planar case, the central frame is determined by $\bar{\theta}$, which is the average of the two particles' rotations. 
Computing the translational shear in this frame removes the need for an additional rotational shear term. 
The generalization of $\bar{\theta}$ is the quaternion spherical linear interpolation.

The equivalence between BPM and FEM requires the shear stiffness $K_s := 12 EI/L^3$ and bending stiffness $K_b := EI/L$. 
The four constants are consistent with the generalized EB beam bond model of \citet{chen_comparative_2022}, whose framework also provides a useful starting point for readers interested in alternative beam models (e.g., Timoshenko).
For a solid cylinder of diameter $d$, $I = \pi d^4/64$ and $J = \pi d^4/32$. 
For a planar beam with cross section $(w, h)$ bending about the $w$ axis, $I = w h^3 /12$. 
When there is no twisting, the spring constants $K_r$, $K_s$, and $K_b$ can be consistently defined even for non-cylindrical bonds.

\subsection{Elastic energy}\label{sec:bpm-energy}

The third expression in Eq.~\eqref{eq:bpm-general-planar} can be obtained via $f_i = -\partial U / \partial v_i$ and $\tau_i = -\partial U / \partial \theta_i$ with the chain rule from the following energy:
\begin{equation}\label{eq:bending-energy}
  U \approx \frac{1}{2} K_s \, \delta s^2
    + \frac{1}{2} K_b \, \theta_b^2.
\end{equation}
Therefore, the total energy of the bond is approximately 
\begin{equation}\label{eq:total-energy}
    U \approx \frac{1}{2} K_r (r_f-r_i)^2 + \frac{1}{2} K_s r_f^2 \gamma^2 + \frac{1}{2} K_t \psi^2 + \frac{1}{2} K_b \theta^2.
\end{equation}

\section{Numerical validation for planar bending}

Since $K_r$ and $K_t$ are straightforward, we validate the remaining parameters $K_s$ and $K_b$ with bending tests on rectangular prismatic beams of dimensions $(L,w,h)$, where $h \ll L$.
Here, $L$ is the total beam length; if a beam is discretized into $N$ bonds, then it is the length $l = L/N$ that enters the BPM parameters in Sec.~\ref{sec:bpm-constants}.
The cross-sectional area is $A = wh$, and we consider bending about the $w$ axis, such that the second moment of area is $I = wh^3/12$.

\subsection{Cantilever beam under static loading}\label{sec:static-beam-convergence}

\begin{figure}[!htpb]
    \centering
    \includegraphics[width=0.9\linewidth]{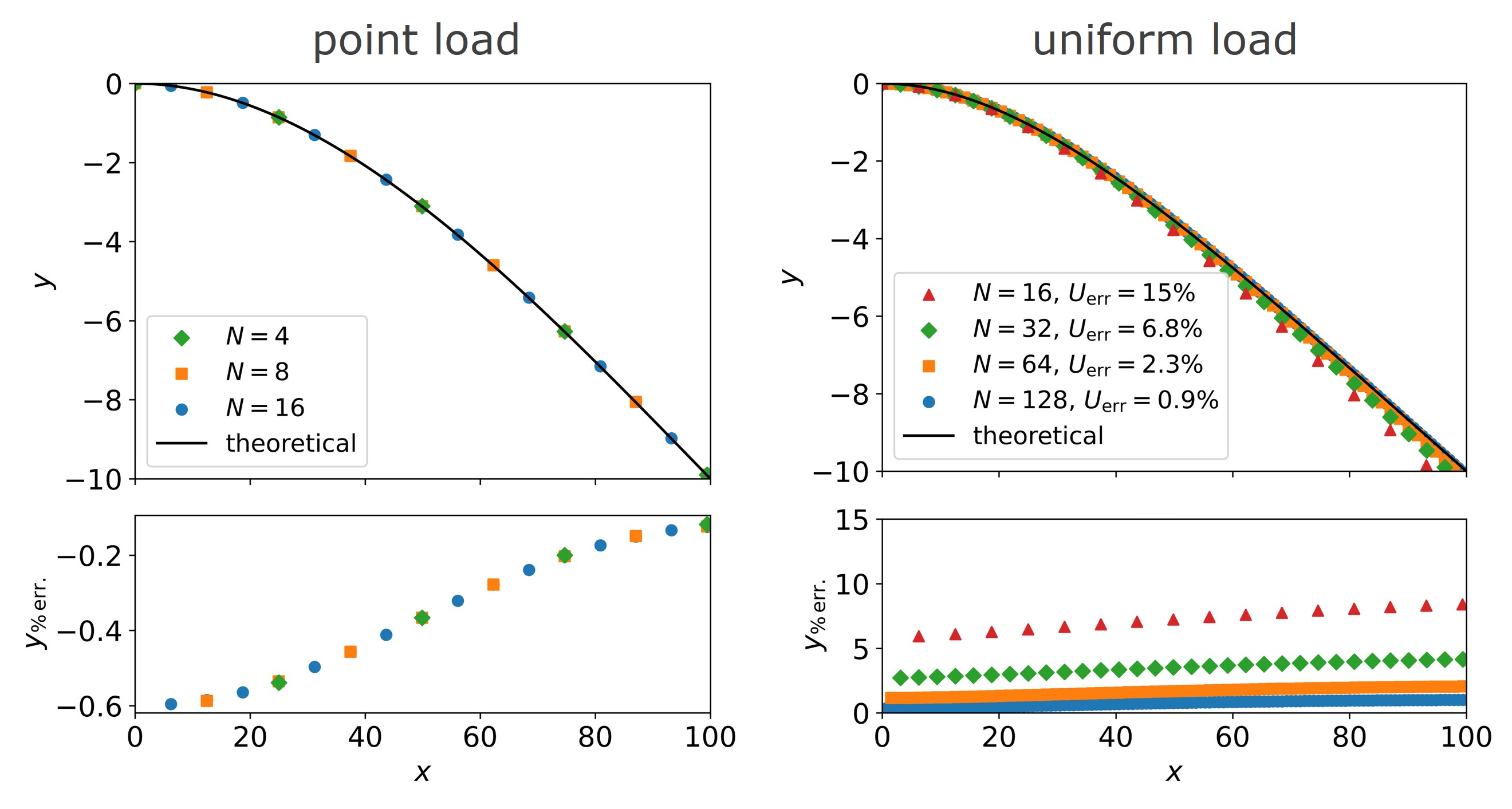}
    \caption{Static bending of a cantilever beam under (left)~a point load and (right)~a uniform load. The simulated deflected shapes for different numbers of bonds $N$ are compared against the exact Euler--Bernoulli solutions (solid curves). The applied loads are chosen such that the maximum tip deflection is $10\%$ of the beam length.}
    \label{fig:planar-static-bending}
\end{figure}

For the first test, we simulated cantilever beams ($L/h = 100$, $E=1$, $\nu=0.5$) under a static 
point load or uniform load. 
The applied loads were chosen such that the maximum tip deflection was $10\%$ of the beam length. 
The simulated shapes are compared against the exact EB deflection curves in Fig.~\ref{fig:planar-static-bending}: 
$y(x) = P x^2 (3L - x) / (6EI)$ for a point load $P$ and 
$y(x) = q x^2 (6L^2 - 4Lx + x^2) / (24EI)$ for a uniform load $q$ \citep{hibbeler2018mechanics}. 
The BPM is accurate even for a small number of bonds under a point load, consistent with the observations of \citet{chen_comparative_2022,guo2013validation,nguyen2013validation}.
However, for a uniformly loaded beam, more bonds are needed to accurately distribute the total load.

We also compared the bond energy (Eq.~\eqref{eq:total-energy}) with the theoretical strain energy at equilibrium: 
$U_\text{theo} = P^2 L^3 / (6EI)$ for the point load and 
$U_\text{theo} = q^2 L^5 / (40EI)$ for the uniform load. 
For the point load, the energy error is approximately $1.5\%$ at $N = 4$ and does not decrease further due to the inherent discretization error. 
For the uniform load, $N \geq 128$ bonds are needed to reduce the error below $1\%$.

\subsection{Cantilever beam under dynamic loading}\label{sec:dynamic-beam-convergence}

\begin{figure}[!htpb]
    \centering
    \includegraphics[width=0.9\linewidth]{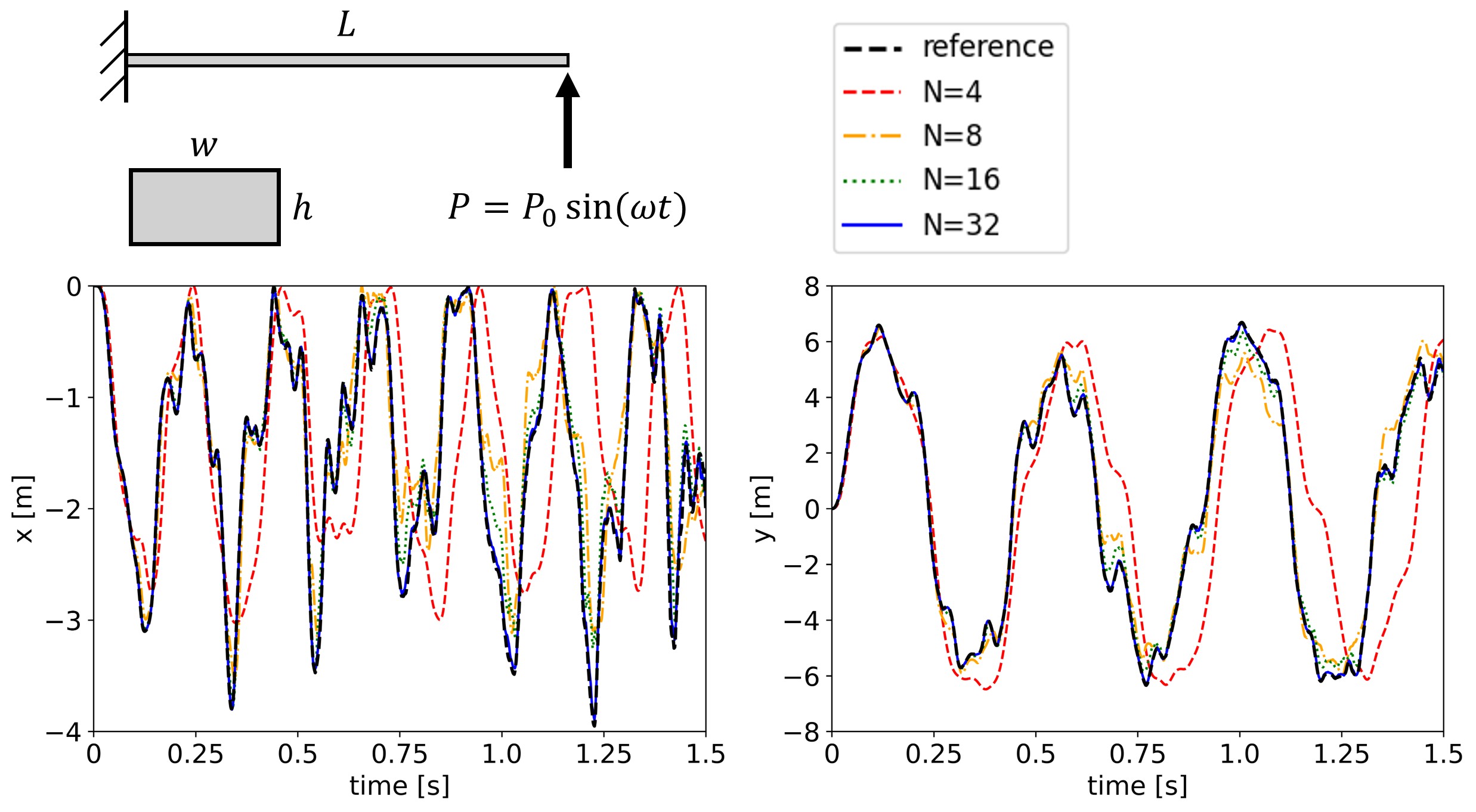}
    \caption{The $x$ and $y$ components of the tip displacement for different numbers of bonds $N$ as a function of time for a cantilever beam under a sinusoidal tip force, compared to the reference solution in \citet{le2011efficient}.}
    \label{fig:planar-dynamic-beam-bending}
\end{figure}

Next, we considered an example in \citet{le2011efficient}. 
A cantilever beam of length $L=10~\unit{m}$, with a uniform rectangular cross section of width $w=0.5~\unit{m}$ and depth $h=0.25~\unit{m}$, has one end clamped and the free end subjected to a sinusoidal tip force $P=P_0\sin(\omega t)$, where $P_0 = 10~\unit{MN}$ and $\omega=50~\unit{rad}/\unit{s}$. 
The beam has an elastic modulus $E=210~\unit{GPa}$ and density $\rho=7850~\unit{kg}/\unit{m}^3$. 
We ran the simulation with no damping, using a time step of $10^{-8}~\unit{s}$. 
The tip displacement is plotted in Fig.~\ref{fig:planar-dynamic-beam-bending} as a function of time for various $N$. 
We observe excellent agreement with \citet{le2011efficient} for $N \geq 32$.

\subsection{Additional data for magnetic beams}\label{sec:magnetic-beam-convergence}

\begin{figure}[!htpb]
    \centering
    \includegraphics[width=\linewidth]{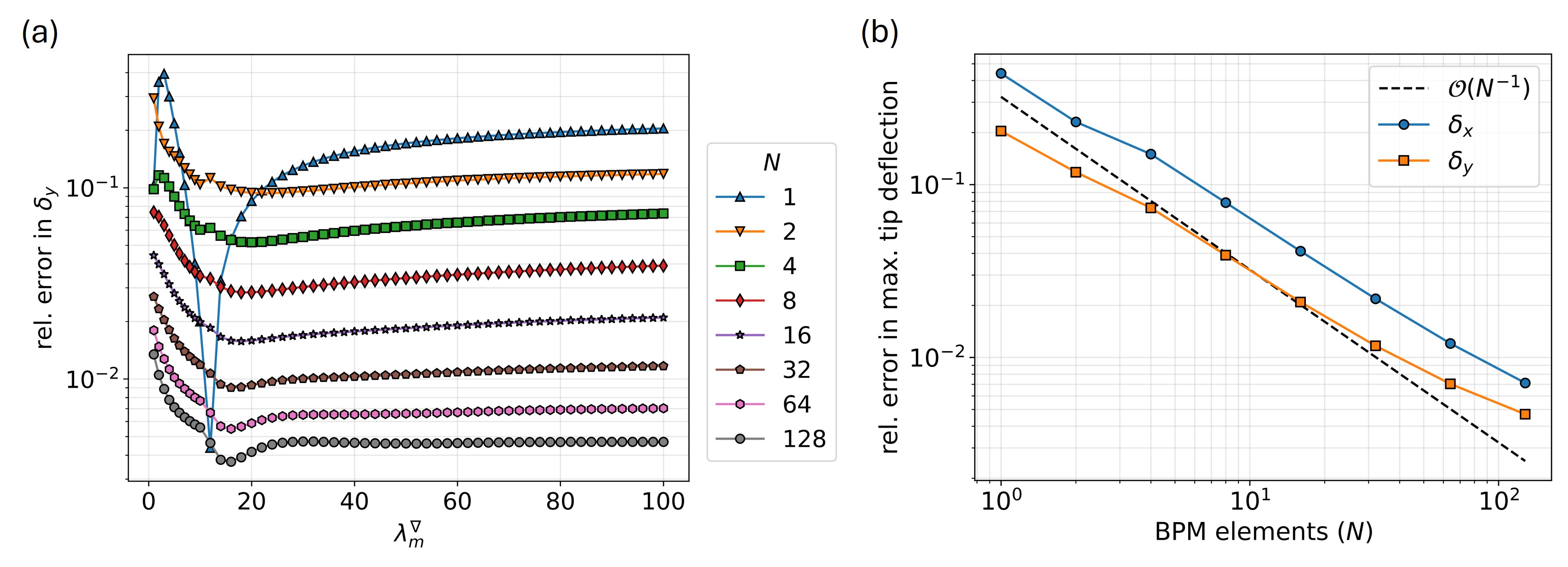}
    \caption{Convergence study for the magnetized beam in Fig.~\ref{M-fig:magnetic-beams}(a) of the main text. (a)~Relative error of the tip deflection in $y$ with respect to the FEM result as a function of the magnetic gradient parameter $\lambda_m^\nabla$ for different numbers of bonds $N$. (b)~Log--log plot of the relative error at $\lambda_m^\nabla = 100$ versus $N$, showing approximately linear convergence.}
    \label{fig:magnetic-beam-convergence}
\end{figure}

We also performed a convergence study for the magnetized beam in Fig.~\ref{M-fig:magnetic-beams}(a) of the main text. 
Fig.~\ref{fig:magnetic-beam-convergence}(a)~shows the relative error of the tip deflection in $y$ with respect to the FEM result as a function of the magnetic gradient parameter $\lambda_m^\nabla$. 
Fig.~\ref{fig:magnetic-beam-convergence}(b)~shows a log--log plot of the relative error at $\lambda_m^\nabla = 100$ as a function of $N$. 
The convergence is approximately linear.

\section{Supplementary videos}

\textbf{Supplementary Video~1.} Side view of the twisting simulation and plectoneme formation of the buckled straight rod (Fig.~\ref{M-fig:heavy-straight-rod}). \\
\textbf{Supplementary Video~2.} Top view of the twisting simulation of the buckled curved rod (Fig.~\ref{M-fig:heavy-curved-rod}). \\
\textbf{Supplementary Video~3.} Magnetic cilia array simulation for $\lambda = L$ (Fig.~\ref{M-fig:magnetic-cilia-array}).

\bibliography{sn-bibliography}